\documentclass[10pt,journal,compsoc]{IEEEtran}

\usepackage[dvipsnames]{xcolor}
\usepackage{amsmath,amsthm,amssymb}
\usepackage{algorithm}
\usepackage{algorithmicx}
\usepackage{bbm,bbold}
\usepackage{ulem}
\usepackage{url}

\usepackage{array}
\usepackage{cite}
\newtheorem{theorem}{Theorem}

\newtheorem{proposition}{Proposition}

\newtheorem{corollary}{Corollary}
\usepackage{subfigure}
\usepackage{color}
\usepackage{graphicx}
\usepackage{flexisym}
\usepackage{dsfont}
\usepackage[dvipsnames]{xcolor}

\definecolor{brown}{rgb}{0.0, 0.0, 0.0}
\definecolor{red}{rgb}{0.0, 0.0, 0.0}
\definecolor{purple}{rgb}{0.0, 0.0, 0.0}
\definecolor{magenta}{rgb}{0.0, 0.0, 0.0}
\definecolor{orange}{rgb}{0.0, 0.0, 0.0}
\definecolor{blue}{rgb}{0.0, 0.0, 0.0}
\definecolor{violet}{rgb}{0.0, 0.0, 0.0}

\usepackage{algpseudocode}

\usepackage{mathtools}

\ifCLASSINFOpdf
\else
\fi

\IEEEoverridecommandlockouts

\begin{document}
%

\title{Towards Effective Resource Procurement in MEC: a Resource Re-selling Framework}

\author{Marie~Siew, \IEEEmembership{Member,~IEEE,} Shikhar~Sharma, Kun~Guo, \IEEEmembership{Member,~IEEE,} Desmond~Cai, Wanli~Wen, \IEEEmembership{Member,~IEEE,} Carlee~Joe-Wong, 
\IEEEmembership{Senior Member,~IEEE, } 
Tony~Q.S.~Quek, \IEEEmembership{Fellow,~IEEE}
\IEEEcompsocitemizethanks{
\IEEEcompsocthanksitem Marie Siew, Shikhar Sharma and Carlee Joe-Wong are with the Electrical and Computer Engineering Department, Carnegie Mellon University, Pittsburgh, PA 15213. \protect\\
E-mails: msiew@andrew.cmu.edu, shikhar2@andrew.cmu.edu, cjoewong@andrew.cmu.edu
\IEEEcompsocthanksitem 
Kun Guo is with the National Mobile Communications Research Laboratory, Southeast University, Nanjing 210096, China, and also with the School of Communications and Electronics Engineering, East China Normal University, Shanghai 200241, China. \protect\\
E-mail: kguo@cee.ecnu.edu.cn
\IEEEcompsocthanksitem Desmond Cai.
E-mail: desmond.cai@gmail.com
\IEEEcompsocthanksitem Wanli Wen is with the School of Microelectronics and Communication Engineering, Chongqing University, Chongqing 400044, China. \protect\\
E-mail: wanli$\_$wen@cqu.edu.cn
\IEEEcompsocthanksitem T. Q. S. Quek is with the Singapore University of Technology and Design, Singapore 487372, and also with the Yonsei Frontier Lab, Yonsei University, South Korea. \protect\\ 
E-mail: tonyquek@sutd.edu.sg
}
\thanks{Corresponding author: Tony Q.S. Quek.}
}

\IEEEtitleabstractindextext{
\begin{abstract}
On-demand and resource reservation pricing models, widely used in cloud computing, are currently used in Multi-Access Edge Computing (MEC). Nevertheless the edge's resources are distributed and each server has lower capacity. If too much resources were reserved in advance, on-demand users may not get their jobs served on time, jeopardizing MEC's latency benefits. Concurrently, reservation plan users may possess un-used quota. Therefore, we propose a sharing platform where reservation plan users can re-sell unused resource quota to on-demand users. To investigate the \textcolor{blue}{mobile network operator's (MNO’s)} incentive of allowing re-selling, we formulate a 3-stage non-cooperative Stackelberg Game and characterize the optimal strategies of buyers and re-sellers. We show that users’ actions give rise to 4 different outcomes at equilibrium, dependent on the prices and supply levels of the sharing and on-demand pools. Based on the 4 possible outcomes, we characterise the MNO’s optimal prices for on-demand users. Numerical results show that having both pools gives the MNO an optimal revenue when the on-demand pool’s supply is low, and unexpectedly, when the MNO’s commission is low. \textcolor{violet}{We develop an interactive prototype, and show that users' decision distributions in studies on our prototype are similar to that of our decision model.}

\end{abstract}

}
\maketitle

%

\section{Introduction}
\label{sec:intro}
\textcolor{violet}{Edge computing enables a wide variety of low latency and computationally intensive services on mobile and other resource constrained devices (e.g Internet of Things (IoT) devices). These low latency services include video analytics, real-time analytics, virtual and augmented reality (VR/ AR), and connected vehicle decision making. Edge computing brings the power of cloud computing to the network edge, with servers placed at edge access points e.g. base stations or wifi access points \cite{PavelMECsurvey}.
Users and device owners can offload computationally intensive tasks to the nearby edge servers, and receive them within latency requirements \cite{mao2017survey}, due to the close proximity of the edge servers.}
The wide-area-network (WAN) delay of cloud computing  will be avoided\cite{hu2015mobile}.

As computing resources are limited at the edge, researchers have been actively investigating the computation offloading and resource allocation problem, 
\cite{mao2017survey, YMaoCompOffloadingJSAC, Xuchen16, dinh2017offloading, li2019resource}. 
Besides the technical challenges of computation offloading, 
\textcolor{purple}{resource pricing 
solutions to optimize revenue \cite{nguyen2019market, TSCmicroeconomic, kim2019economics, liICASSP19, wang2021eihdp, wang2019profit, siew2021let, halima2017optimal, ben2021optimal} can serve as complementary mechanisms to control user demands 
and ensure good quality-of-service despite limited available resources}, \textcolor{magenta}{while maintaining the network operator's incentive.}
\textcolor{magenta}{Currently, Edge computing providers have been adopting the traditional cloud computing pricing models \cite{ec2Pricing,azurePricing, wang2012cloud}: the \textit{On-Demand pricing} model (OPM) and the \textit{Resource Reservation} model (RRM), e.g. in AWS Local Zones \cite{edge-AWSlocalZones} and Juniper \cite{edge-Juniper}.} 
These two models cater to different types of users. 
In the \textit{On-Demand pricing} model, service users pay for computing resources, as and when they need it, 
\textcolor{black}{and do not make any long term commitment \cite{on-demandAmazon}}. 
\textcolor{violet}{In MEC, this is suitable for individual mobile users using, for example, virtual reality or mobile gaming services once, or a couple of times, 
not across consistent timeslots across the span of months;} for firms during special one-off events involving augmented reality displays or machine learning platforms. 
In contrast, for the \textit{Resource reservation} model, \textcolor{violet}{service
users reserve computing resource instances in advance,} across multiple timeslots, \textcolor{brown}{(e.g. for a year),} 
for a discounted price \cite{amazonRI}. This caters to users (or firms) who use computing resources in bulk, at periodic and pre-determined timings. These may include IoT vendors with periodic data analytic requirements for their IoT networks or firms behind public augmented reality setups, with a constant need of computing capabilities.


\textbf{\textcolor{magenta}{Besides maximizing revenue, will such plans be able to 
meet user demand?}} 
%
\textcolor{brown}{While edge computing is advantageous latency-wise through the proximity to end users,} unlike cloud data centers with thousands of servers \cite{EClatencyComp}, 
the distributed nature of MEC involves small-scale data centers placed at the network edge \cite{mao2017survey, PavelMECsurvey}. The server capacity at edge nodes may be limited \textcolor{brown}{due to physical constraints \cite{li2019resource}.} 
Therefore, the difficulty from jointly using the \textit{On-Demand pricing} and \textit{Resource Reservation pricing} models in MEC is that after the resource units are reserved in advance by the \textit{Resource Reservation} users, there \textbf{might be fewer resource units left for on-demand users}.
\textcolor{violet}{Consequently, on-demand users may experience a longer service latency, which may jeopardize the utility of MEC}
\textcolor{violet}{since services which use edge computing are often latency sensitive 
\cite{mao2017survey,hu2015mobile}.} 
\textcolor{brown}{Cloud providers complement the on-demand and resource reservation models and \textcolor{purple}{monetize} excess demand through spot instances, leasing the unused capacity at data centers to users at a discounted price \cite{spotInstances}.
We are unable to consider spot pricing in MEC because unlike the cloud, there is a lack of spare capacity at the edge.}
\textcolor{purple}{
Furthermore, raising the price of resource reservations to control the demand, may not suffice as users with long-term usage needs may continue to prefer fixed upfront expenses compared to on-demand expenses that may fluctuate and may be hard to predict.}

\textcolor{violet}{Therefore, to improve the QoS of users, we propose a sharing platform, titled the \textit{\textbf{Sharing Quota Model}} (SQM),} 
Here, \textcolor{black}{reservation plan users can re-sell their excess un-utilized resource quota via the platform to on-demand users, while the MNO collects a commission.}
{\color{black}This could benefit all parties. 
The on-demand users can attain a higher QoS \textcolor{black}{(lower delay)} and a higher probability of getting their jobs served, the reservation plan users will obtain extra income 
and the MNO collects a commission.} Higher customer satisfaction also benefits the MNO in the long run. 
\textcolor{orange}{It is unclear, however, whether the MNO always has an incentive to offer such a platform, as the SQM may discourage on-demand users from purchasing capacity directly from the MNO.}
\textcolor{black}{To best integrate} the sharing platform with the on-demand pricing and resource reservation 
models, the following \textcolor{purple}{research} questions arise:

\textit{a) Given the sharing platform, would on-demand users still buy from {\color{black}the on-demand platform}? What would the aggregate equilibrium behavior 
look like?}

\textit{b) Given the users' decisions, \textcolor{violet}{it is not clear if the MNO has an incentive to provide this platform: will it lead to sub-optimal revenue?} 
How should it set prices such that revenue is optimized, \textcolor{black}{whilst minimizing the unmet demand of the on-demand users?}} 

\textcolor{orange}{\textit{c) How can we design the platform? Is it more practical if resource quota re-selling is automatically or manually decided? }}

\textcolor{violet}{While there have been resource sharing (or trading) platforms in networks, e.g. mobile data market trading platforms \cite{mobileDataTrading1,mobileDataTrading2, mobileDataTrading3}, and crowdsourced community wireless networks \cite{wifiSharing}, their market structures are different. Specifically, the mobile data market users buy plans which have the same structure, differentiated by usage levels. In crowdsourced community wireless network model formulations, given prices and membership choices of other users, users decide their network access time. This differs from our market structure, where there are two types of plans (on-demand vs reservation), determined by service application type and usage patterns.
Our work also differs from other MEC resource allocation or revenue maximization work, as it uniquely considers how the on demand and resource reservation model can be improved in MEC.}
Therefore, to answer the above questions \textcolor{magenta}{and investigate the feasibility of the sharing platform, we perform game theoretic analysis on the three-sided market, formulating a three-{\color{black} stage} {\color{black} non-cooperative} Stackelberg Game amongst the MNO,} {\color{black}re-}sellers {\color{black}(reservation plan users) and buyers (on-demand users).} 
Our \textbf{analysis characterises the decision making and conflicting incentives of the three parties}, with the MNO setting the on-demand pool and sharing pool's prices to optimize revenue, the on-demand users choosing the pool to purchase from,  
aiming to maximize their payoff functions, 
and the reservation users also aiming to maximize their payoff functions. 

Our contributions are summarized as follows:
\begin{itemize}
    \item \textcolor{blue}{
    With a direct application of on-demand and resource reservation pricing plans in MEC, on-demand users face potentially limited resource availability at the edge, \textcolor{violet}{impacting their QoS,} while resource reservation plan users might have reserved excess un-utilized resource quota.
    Therefore we propose a novel model integrating a sharing platform, for the re-selling of unused resource quota from reservation plan users to on-demand users.} 
    \item \textcolor{blue}{To investigate the MNO's incentive on having a sharing platform and to optimize the MNO's prices for revenue maximization, we will use a \textbf{3-stage Stackelberg Game formulation} to
    characterise how the MNO, buyers and re-sellers interact at equilibrium.} 
    In this sequential game, the MNO selects prices to optimize its revenue. \textcolor{magenta}{Following which, the re-sellers (reservation plan users) choose whether or not to sell, forming the sharing pool. Viewing the prices and relative supplies, the buyers choose which pool (or not) to buy from.} We incorporate two platform design options - automated re-selling of unused quota, and manual decision making for re-sellers. 
    \item 
    \textcolor{blue}{Using backwards induction, we firstly \textbf{characterise the buyer's optimal strategies} for the above two platform design options (Proposition \ref{prop:buyersStrategies}). 
    Our analysis shows that the \textbf{equilibrium behavior amongst the buyers and re-sellers
    possesses a 4 region structure} (Proposition \ref{prop:4regions}, with \textbf{each region corresponding to whether or not the on-demand and sharing pools coexist}). The likelihood of each of the 4 regions happening depend on the relative magnitudes of prices and supply levels of the two pools.}
    Based on the 4 region structure, we optimize the on-demand and sharing prices (which also impacts the sharing supply), to \textbf{optimize the MNO's revenue, and provide insights on the optimal price} \textcolor{red}{(Theorems \ref{thr:buyerGameOpt} and \ref{thr:optimalPo}, Corollary \ref{corollary:OptRev}).} 
    \item We provide numerical {\color{black}results}, which show how the system behaves at the optimal point. Our results show that \textcolor{blue}{for different willingness to pay and usage distributions, having both the sharing and on-demand pool gives the MNO an optimal revenue 
    when the on-demand pool's supply $q_o$ is low,} and when $\delta$, the platform's commission is low,
    but not below a certain threshold. It would be expected that a high commission would give the platform a higher sharing revenue, nevertheless our results show that it incurred a trade-off in dissuading reservation plan users from re-selling.
    \item \textcolor{violet}{We created \textbf{a prototype, an interactive website, to conduct preliminary user studies on user decision making} in our resource re-selling framework.
    Our analysis shows that there is no statistically significant evidence that the empirical distribution function of the \textit{data from the user study} and the empirical distribution function of the \textit{data from the model's predictions} (Eq. 1, Prop. 1-2) are distinct.
    } 
\end{itemize}

The rest of this paper is organized as follows. In Section \ref{sec:related}, we introduce the related works. In Section \ref{sec:model}, we describe the system model. Following which, we {\color{black}formulate the Stackelberg Game in Section} \ref{sec:ProbForm}. In Section \ref{sec:gameAnalysis}, we present the game analysis and insights. We present numerical results in Section \ref{sec:simulations}, present our prototype and preliminary users studies in Section \ref{sec:prototype}, discuss future areas of investigation in Section \ref{sec:futurework}, and finally conclude this paper in Section \ref{sec:conclusion}.

\section{Related Works}
\label{sec:related}
\textcolor{violet}{In this work, we propose a novel system model incorporating a resource re-selling platform, to improve the integration of the on-demand and resource reservation pricing models in MEC, given the limited resources at the edge. We also discuss how our model differs from other platforms in networked systems.}

\textcolor{blue}{\textbf{Edge Computing resource allocation:}} Researchers have been actively 
\textcolor{red}{optimizing resource allocation in MEC,} 
balancing the user's delay requirements with minimizing the energy consumption, under various scenarios \cite{YMaoCompOffloadingJSAC, Xuchen16, dinh2017offloading, li2019resource, zhao2022ctl}. 
Another line of work is collaborative edge computing to maintain the QoS in light of user mobility. This involves strategies like VM or service migration across base stations \cite{ouyang2018follow, ma2020leveraging, wang2019delay, siew2021let} and user association optimization \cite{dai2018joint}. 

\textcolor{blue}{\textbf{Edge Computing revenue maximization:}} 
These works involve optimizing revenue while maintaining the quality of service for users (delay, job service rate, etc) \cite{nguyen2019market, TSCmicroeconomic, kim2019economics, wang2021eihdp, wang2019profit,liICASSP19}. 
For example, \cite{nguyen2019market} used the Fisher market model to model the MEC market. 
\cite{TSCmicroeconomic} jointly optimized over the resource prices and the devices' budget allocation strategies. 
\cite{kim2019economics} modelled a dynamic game 
between the edge resource owners, infrastructure providers, and service users. 
\textcolor{brown}{While our model follows the user payoff model of \cite{kim2019economics}, we model a distinct scenario, involving a novel hybrid sharing, on-demand and resource reservation framework. \textcolor{purple}{Thus, unlike~\cite{kim2019economics} we must further account for re-seller incentives, leading to distinct results. }}
\textcolor{blue}{In our prior work \cite{siew2020dynamic}, we proposed a model where users constantly change whether to be an 'owner' or 'renter' of excess resource units, along with dynamic pricing mechanisms.
We did not investigate the MNO's incentive of having a sharing platform. To the best of our knowledge, there has not been work on how to improve the integration of the on-demand and the resource reservation pricing models in MEC} 

\textcolor{blue}{\textbf{Cloud Computing pricing:}} Maximizing revenue in cloud computing by improving on the on-demand and reserved instance models was studied in \cite{wang2012cloud,xu2013dynamic, wang2012towards, alzhouri2018maximizing}.
These works focused on spot instances, which sell the spare capacity at data centers, at discounted prices \cite{spotInstances}.
For example, \cite{wang2012cloud} proposed a truthful auction for spot instances. 
\cite{xu2013dynamic} proposed a dynamic pricing policy. 
\cite{wang2012towards} maximized revenue through capacity segmentation. 
Unlike in cloud computing, each edge server has limited resources, making the spot instances model not feasible in edge networks. 

\textcolor{blue}{\textbf{Other Platforms in Networked Systems:}} \textcolor{brown}{The mobile data market has seen the existence of platforms (e.g. China Mobile Hong Kong’s 2nd exChange
Market) which provide a secondary data market, allowing users to buy and sell data from each other \cite{mobileDataTrading1,mobileDataTrading2}. 
These works \cite{mobileDataTrading1,mobileDataTrading2, mobileDataTrading3} perform a game theoretic analysis to characterise the equilibrium and optimal decisions under various scenarios, as well as propose alternative platform designs \cite{mobileDataTrading4}.
\textcolor{blue}{This platform has a different market structure from the edge computing market} - here all the users buy resources under the same type of scheme (monthly data plans with usage quotas) \textcolor{blue}{and are differentiated by usage levels. In the edge computing market, users have vastly different kinds of applications and usage patterns, which suit different usage schemes: some users have consistent usage patterns (e.g. IoT vendors) and suit the resource reservation model, and some users have more sporadic, one-off usage patterns (e.g. individual mobile users using applications like virtual reality) and are suited to the on-demand model. 
}  
}


\section{The Resource Procurement Model}
\label{sec:model}
We consider an edge MNO that provides edge computing service to $N$ users.
The MNO has 2 resource procurement schemes for users to choose from, namely the \textit{\textbf{Resource Reservation Model}} (RRM) and \textit{\textbf{On-Demand Pricing Model}} (OPM). 
\textcolor{magenta}{These models, which are commonly used as cloud pricing and procurement schemes 
\cite{ec2Pricing, azurePricing, wang2012cloud}, are currently being used in edge networks \cite{edge-AWSlocalZones,edge-Juniper}.} 
Under OPM, the users purchase credits to offload their jobs at the edge servers "on the go", as and when they need it. They pay $p_o$ per unit of resource consumed (CPU cycles), to the MNO. Customers of on-demand pricing could include individual mobile users, e.g. users of applications like AR/VR or mobile gaming, or firms who are using AR or machine learning applications for one-off special events or displays. These users' demand for computing is sporadic, not following a consistent usage pattern across timeslots.

The RRM is catered to customers who have consistent computing requirements across different timeslots. The RRM allows them to reserve \textcolor{brown}{(i.e. commit to) computing resource in advance across multiple timeslots, e.g. for a year, at a discounted price \textcolor{purple}{per timeslot} of $p_r$.} 
Customers of the reservation plans include users who use computing resources in bulk, at periodic timings, and more frequently in general.
For example, \textcolor{purple}{these may include} IoT vendors with constant sensing and data analysis requirements for their large IoT network, autonomous driving firms, entities behind long-term public augmented reality setups. 

The user has chosen between the resource {\color{black}procurement} schemes OPM and RRM based on its \textcolor{violet}{service application's usage requirements:} sporadic, one-off vs consistent, frequent and long-term.
\textcolor{brown}{The details of the user usage model will be in covered the following subsections.}
\textcolor{black}{The computing resources at the edge server are first provisioned to the reserved users, with the remaining resources forming the on-demand pool.} 

\begin{figure}[t]
\centering
\includegraphics[
angle=0,scale=0.2]{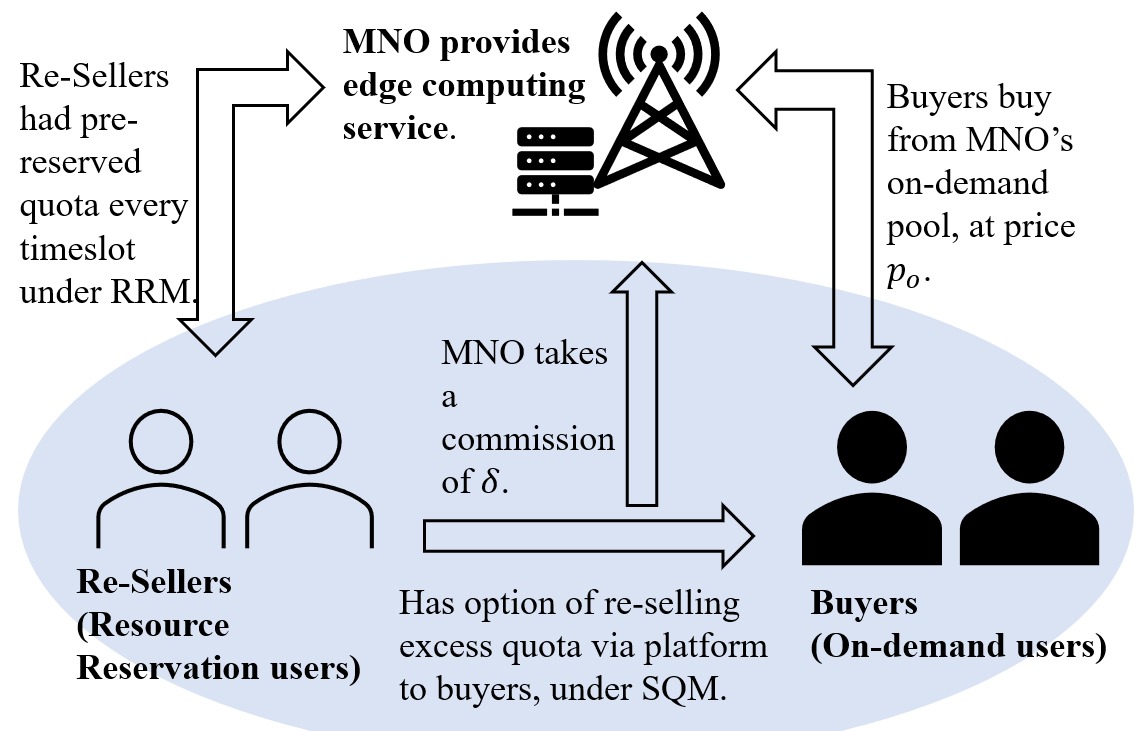}
\caption{The integration of the resource-quota sharing platform with the On-Demand and Resource Reservation models.}
\label{fig:sysModel}
\end{figure}

\textcolor{black}{Nevertheless, unlike cloud resource allocation where data centers have huge amount of computing resources, the server capacity at edge nodes is more limited \cite{mao2017survey,PavelMECsurvey}, and the data processing rate is less powerful \cite{ECtaskProcessingRate}.}
Hence, after resources are reserved in advance by RRM users, there might be less resources left for on-demand users.
At the same time,
RRM users may find that for specific timeslots, they have excess computing resource quota for CPU cycles which they do not require.
In this work we propose and explore the impact of adding a resource sharing scheme, the \textit{\textbf{Sharing Quota Model}} (SQM). 
In this integrated scheme (see Figure \ref{fig:sysModel}), RRM users act as \textit{re-sellers}, selling their excess unused resource quota at a price of $p_r$ {\color{black}per unit of resource} to on-demand users (\textit{buyers}).
\textcolor{purple}{The resold resources} \textcolor{black}{form the sharing pool and} increase the total supply available to on-demand users (\textit{buyers}).
The MNO acts as a middleman and takes a commission of $\delta$ from the transaction. 
We investigate the impact that adding SQM has on the profits of the 
MNO, and hence whether this scheme is feasible.

\setlength{\textfloatsep} {3pt}



\textit{\textbf{Re-sellers:}} The \textit{re-sellers} are users who have previously reserved resources from \textbf{RRM}, and who have excess resources available for sale. Under the proposed sharing scheme \textbf{SQM}, they have 2 choices $s=\{sell,no\}$ \textcolor{purple}{in each timeslot}. 
They can choose whether to sell their excess computing resource quota via the MNO's platform to on-demand users (\textit{buyers}), or not. {\color{black}Without loss of generality, we consider the payoff per unit resource quota for seller $i$, which is expressed as:}
\begin{equation}
    \label{sellersPayoffEqns}
    \pi_i^s(g_i)= \begin{cases} (1-\delta)p_r -g_i, & \text{if} \ s=sell\\
    0, & \text{if} \ s=no.
    \end{cases}
\end{equation}

{\color{black}Recall that $p_r$ is the per unit price set by the MNO for the transaction, and define $\delta$ as the commission level which the MNO takes from the transaction.
} We let $g_i$ represent the inconvenience cost of selling for seller $i$ (factoring in uncertainty of usage, and the inconvenience of checking the app to decide to share). 
\textcolor{brown}{In this paper, we consider per unit resources in the payoff functions of both buyers and re-sellers. Buyers and re-sellers with multiple units are considered as multiple buyers or re-sellers.
Note that if we had considered users buying/re-selling distinct quantities of resources, with the resource usage level of users following $U[0,1]$, the impact would be that the supply and revenue functions are multiplied by a constant factor.}

\textcolor{black}{The inconvenience cost depends on the extent to which each re-seller 
is willing and able to expend time, energy and resources to periodically check the app, factor in the uncertainty of personal usage, and decide whether or not to sell the excess resource quota.} 
\textcolor{black}{This differs across different vendors and owners, and differs across applications.} 
\textcolor{magenta}{For example, it would be easier for an IoT vendor to dedicate resources to check the app, as comapared to an individual MEC user. Some individual MEC users are more likely to check than others as they are more interested in earning extra income. Besides this ,it is more difficult for users of some applications to make the decision to re-sell, as their usage for that particular timeslot is yet to be known, and more unpredictable.}
$g_i$ is a random variable which follows the uniform $U[0,1]$ distribution. It captures the varying levels of personal costs across the users. 
\textcolor{brown}{
While the distribution has an impact on the way revenue is computed, we perform simulations using both uniform and beta (which has a gaussian-like structure) distributions in Section \ref{sec:simulations}, showing similar results across distributions.}

As we can see from Eq. (\ref{sellersPayoffEqns}),
\textcolor{brown}{a re-seller's decision is impacted by the price it receives $p_r$, the MNO's commission level $\delta$, and the inconvenience cost $g_i$.} 
Re-seller $i$ will share/sell when its profit $(1-\delta)p_r$ is greater than its cost $g_i$. If the inconvenience cost is too high, it will not.
To obtain the total proportion of re-sellers willing to sell/share, we integrate over the users whose inconvenience cost $g_i$ is less than or equals to $(1-\delta)p_r$.
As $g_i$ follows the uniform $U[0,1]$ distribution, the proportion of re-sellers willing to share/sell will be:
\begin{equation}
    \begin{split}
    prop & =  \int^{(1-\delta)p_r}_0  d g_i\\
    & = (1-\delta)p_r.
    \end{split}
    \label{propEqn}
\end{equation}
{\color{black}}

\textcolor{brown}{The \textit{quality} of the sharing pool, $q_s$, indicates the total supply of resource quota available for buyers, and therefore is dependent on the 
proportion of re-sellers who are willing to share their unused quota, $(1-\delta)p_r$.}
It can be expressed as\footnote{{\color{black}Function $q_s(p_r)$ may have various forms in realistic MEC systems, \textcolor{purple}{but our analysis permits any functional form. }
We will discuss a concrete form for numerical results in Section \ref{sec:simulations}.}}:
\begin{equation}
    q_s(p_r)=f((1-\delta)p_r),
    \label{eq:supplyGeneric}
\end{equation}
\textcolor{brown}{with $q_s(p_r)$ being a strictly increasing, twice differentiable and concave function.}
\textcolor{violet}{This supply level has an impact on the user experienced delay.}


\textit{\textbf{Buyers:}} The \textit{buyers} are on-demand users who purchase quota to offload their
jobs at the edge servers on the go, i.e. not in advance. 
These buyers send their computing jobs to the edge server, with different willingness to pay (represented by $u_i$ for user $i$). The willingness to pay is the buyer's utility of job computation at the edge server.
For example, some jobs are more urgent and considered more of a necessity than others (e.g. public AR/VR displays, autonomous vehicle computation), and therefore the buyers will still be willing to pay despite the price $p_r$ increasing. 
On the other hand, some jobs are considered less urgent, or are not viewed as necessities from the buyer's perspective, resulting in a lower willingness to pay: given a high price of resource quota $p_r$, the buyer is more likely to choose not to purchase the edge computing service.
We let $u_i$ follow the uniform $U[0,1]$ distribution, to represent how the willingness to pay differs across users.
\textcolor{brown}{Note that the distribution has no impact on the payoff maximizing strategies of the buyers, or the equilibrium outcome of user decisions in light of fixed prices. While it has an impact on the way revenue \textcolor{red}{(Eq. (\ref{eq:revenue_MNO}))} is computed, 
we perform simulations using both uniform and beta distributions in Section \ref{sec:simulations} which show similar trends and effects across distributions.}

Under \textbf{SQM}, these on-demand users now have 3 options $b\in \{\text{on-demand pool, sharing pool, {\color{black}none}}\}$.\footnote{{\color{black}For simplification, we use $b=o$, $b=s$, and $b=n$ to represent $b=\text{on-demand pool}$, $b=\text{sharing pool}$, and $b=\text{{\color{black}none}}$, respectively.}} {\color{black}Note that, $b=\text{none}$ means not buying from both the on-demand pool and sharing pool. Therefore} 
the payoff of buyer $i$ is:
\begin{equation}
\label{eq:BuyerPayoff}
    \pi_i^b(u_i)= \begin{cases} u_i q_o-p_o, & \text{if} \ b= \text{on-demand pool}\\
    u_i {\color{black}q_s(p_r)}-p_r, & \text{if} \ b= \text{sharing pool} \\
    0, & \text{if}\ b= \text{not buying},
    \end{cases}
\end{equation}
where $u_i$ is buyer $i$'s willingness to pay for a unit resource quota, as described earlier \textcolor{purple}{and used in~\cite{kim2019economics}}. 
The payoff when users choose the on-demand and sharing pool respectively is their willingness to pay (satisfaction) $u_i$ multiplied by the quality/supply of the pool ($q_o$ or $q_s$) minus the payment ($p_o$ or $p_r$). 
The quality/supply ($q_o$ or $q_s$) \textcolor{brown}{indicates the relative supply of resources available at the on-demand and sharing pool respectively. 
The quality at the on-demand pool, $q_o$, is the amount of supply of computing resources (in CPU cycles) at the on-demand pool, after the resource reservation users have reserved their portion.\footnote{\textcolor{black}{Note that in traditional centralised cloud computing, the supply at data centers (i.e. the $q_o$ equivalent) would be \textcolor{red}{much larger in magnitude,} \textcolor{magenta}{resulting in less of a need for a sharing pool, according to Eq. (\ref{eq:BuyerPayoff}).}}} 
As mentioned earlier, the quality at the sharing pool, $q_s(p_r)$, is a function of the proportion of re-sellers who are willing to re-sell.
\textcolor{violet}{The higher the supply at each pool, the higher the likelihood the buyer is able to get its job served, hence improving its QoS in terms of delay experienced.}
Therefore the buyer's payoff $\pi_i^b(u_i)$ is an increasing function of the quality $q_s(p_r)$ or $q_o$.}
As different buyers have different willingness to pay $u_i$, different buyers will make different choices, based on which choice maximizes their individual payoff. 
\textcolor{magenta}{
In the next few sections, we investigate the equilibrium behavior of the users, and the optimal decisions for the MNO to make based on that.}
{\color{black}We summarize the key notations used throughout this paper in Table \ref{tab:notations}}.


\begin{table}[t]
%
\centering
\caption {Summary of key Notations}\label{tab:notations} 
 \begin{tabular}{|c|c|} 
 \hline
 \textbf{{\color{black}Notation}} & \textbf{Definition} \\ 
 \hline
 $u_i$ & Willingness to pay of buyer $i$.\\
 \hline
 $g_i$ & The inconvenience cost of re-seller $i$.\\
 \hline 
 $\delta$ & The MNO's commission percentage. \\
 \hline
 $p_r$ & Price per-unit of re-sold quota at the sharing platform. \\ 
 \hline
 $p_o$ & Price per-unit resource at the on-demand pool. \\
 \hline
 $q_o$ & Supply, and hence quality at the on-demand pool.\\
 \hline
 {\color{black}$q_s(p_r)$} & Supply and hence quality at the sharing platform.\\
 \hline
 $\pi_i^s(g_i)$ & Payoff of re-seller $i$.\\
 \hline
 $\pi_i^b(u_i)$ & Payoff of buyer $i$.\\
 \hline
 $R(p_r,p_o)$ & Revenue of the MNO. \\ 
 \hline
\end{tabular} 
\end{table}

\section{Problem Formulation}
\label{sec:ProbForm}
The MNO's objective is to optimize its revenue, when integrating the sharing {\color{black}scheme} \textbf{SQM} with the existing pricing models.
\textcolor{violet}{The MNO's decisions will be made based on the aggregate behavior of behavior of buyers and re-sellers at equilibrium.}
Its Revenue Maximization Problem (\textbf{RMP}) is:
\begin{equation}
\textbf{RMP}: \underset{p_o,p_r}{\text{max}}\ R(p_r,p_o)
\end{equation}
where 
\begin{equation}
\begin{split}
    R(p_r,p_o) = & N \int^1_0 p_o  \mathbb{1}_{\{\pi^{b=o}_i(u_i)>max(\pi^{b=s}_i(u_i),0)\}} d u_i \\
    & + N \int^1_0 p_r \delta \mathbb{1}_{\{\pi^{b=s}_i(u_i)>max(\pi^{b=o}_i(u_i),0)\}} d u_i.
\end{split}
\label{eq:revenue_MNO}
\end{equation}
{\color{black}In Eq. (\ref{eq:revenue_MNO}), $N$ is the number of users, and $p_o$ and $p_r$ are the price per unit resource at the on-demand pool and the sharing pool, respectively. Hence, the first term on the right hand side is the MNO's} revenue attained from the on-demand pool, where $\mathbb{1}_{\{\pi^{b=o}_i(u_i)>max(\pi^{b=s}_i(u_i),0)\}}$ is the indicator function indicating if buyer $i$ gains a higher payoff from choosing the on-demand pool, over the sharing pool and not buying at all.
The second term is the MNO's revenue from the sharing pool, where $\mathbb{1}_{\{\pi^{b=s}_i(u_i)>max(\pi^{b=o}_i(u_i),0)\}}$ is the indicator function indicating if buyer $i$ gains a higher payoff from choosing the sharing pool, over the on-demand pool and not buying at all.

\begin{figure}[t]
\centering
\includegraphics[angle=0,scale=0.4]{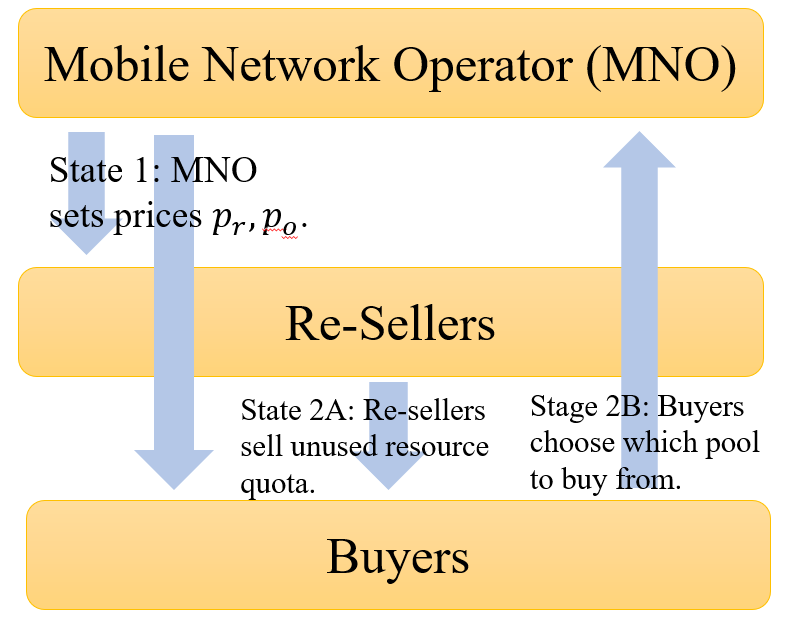}
\caption{\textcolor{black}{Game formulation: Non-cooperative game between the MNO, re-sellers and buyers.} 
}
\label{fig:sequenceEvents}
\end{figure}

To analyse the strategic behaviours \textcolor{black}{and} the dynamic interaction amongst the three parties and to obtain equilibrium insights, we \textcolor{black}{formulate a non-cooperative Stackelberg Game \cite{mas1995microeconomic} (Fig. \ref{fig:sequenceEvents}). }
In Stage 1, the MNO (the leader) sets the on-demand and sharing pool prices $p_o$ and $p_r$, to maximize its revenue $R(p_r,p_o)$, {\color{black}as in Eq. (\ref{LeaderEq})}.
In Stage 2A, given $p_r$ and $\delta$, the \textit{re-sellers} (followers) individually decide $s$, whether they would like to share or not. A re-seller's decision depends on which choice maximizes their individual payoff $\pi_i^s$, {\color{black}following Eq. (\ref{Follower1Eq})}. The collective decision of all re-sellers would result in a supply/quality of $q_s(p_r)$ for the sharing pool, which would influence the decisions of the buyers in Stage 2B.
In Stage 2B, given the prices $p_r$ and $p_o$, and the supply/qualities $q_s(p_r)$ and $q_o$, the \textit{buyers} (followers) individually decide $b$, which pool they would buy from.
A buyer's decision depends on which choice maximizes their individual payoff $\pi_i^b$, {\color{black}as presented in Eq. (\ref{Follower2Eq})}.
The decisions of the followers (the re-sellers and buyers) would in turn affect the revenue of the {\color{black}leader} MNO.
\setlength{\textfloatsep} {3pt}
\begin{equation}
\label{LeaderEq}
Stage \ 1: \text{MNO (leader): } (p_r^*,p_o^*)=\underset{p_r,p_o}{\text{argmax}} R(p_r,p_o)
\end{equation}
\begin{equation}
\label{Follower1Eq}
Stage \ 2A: \text{Re-sellers:} \ s^*=\underset{s \in\{s,n\}}{\text{argmax}} \ \pi_i^s(g_i)\\
\end{equation}
\begin{equation}
\label{Follower2Eq}
Stage \ 2B: \text{Buyers :} \ b^*=\underset{b \in\{o,s,n\}}{\text{argmax}} \ \pi_i^b(u_i)  
\end{equation}


\section{Game Analysis}
\label{sec:gameAnalysis}
In this section, we solve the Stackleberg Game (Eqs. (\ref{LeaderEq}), (\ref{Follower1Eq}) and (\ref{Follower2Eq})), 
\textcolor{violet}{characterizing and presenting insights on the equilibrium interaction amongst the MNO, buyers, and re-sellers, and characterizing the MNO's optimal prices.}
\textcolor{black}{We analyze two distinct platform designs which the MNO could implement for the re-selling portion of the sharing platform.}
Firstly, in {\color{black} Section} \ref{subsec:MNObuyers}, we explore the option
\textcolor{black}{where the MNO implements automatic re-selling of unused resource quota on the part of users. While this may not be as feasible in practice \textcolor{magenta}{as it involves all users knowing upfront whether or not they have excess quota,}
\textcolor{brown}{nevertheless, this portion of the study gives insights on buyer decisions and the aggregate outcome at equilibrium in light of MNO prices, by isolating the MNO-buyer interaction.}}
In {\color{black}Section} \ref{subsec:combined} we \textcolor{black}{explore the option where the MNO implements manual decision making for re-sellers: re-sellers weigh their personal inconvenience cost $g_i$ when deciding whether or not to re-sell quota in the sharing pool, according to Eq. (\ref{sellersPayoffEqns}).} 

\subsection{Case 1: Automatic Re-Selling of Unused Quota}
\label{subsec:MNObuyers}
In this subsection we analyze the scenario where the \textcolor{black}{MNO implements automatic re-selling of unused resource quota on the part of re-sellers.} 

\textcolor{black}{As there is no decision making on the part of re-sellers, we have $q_s(p_r)$ as a constant $q_s$.}
The resulting game between the MNO and the \textit{buyers} is: 
\begin{equation}
Stage\ 1: \text{MNO decides} \ (p_r^*,p_o^*)=\underset{p_r,p_o}{\text{argmax}}\ R(p_r,p_o)
\end{equation}
\begin{equation}
Stage\ 2: \text{Buyers decide} \ b^*=\underset{b \in\{o,s,n\}}{\text{argmax}} \ \pi_i^b(u_i)
\end{equation}

\subsubsection{The buyer's optimal strategies}

We solve this game by backward induction \cite{mas1995microeconomic}, solving Stage $2$ first, and using the equilibrium outcome of Stage $2$ to solve Stage $1$ and obtain the MNO's optimal prices.

Buyers try to maximize their payoffs $\pi_i^b$ by choosing whether they will buy computing resource quota from the on-demand pool, the sharing pool, or not buy at all, given the prices $p_o$, $p_r$, and the relative supply $q_o$ and $q_s$. Proposition \ref{prop:buyersStrategies} shows which strategies {\color{black} buyer $i$} will take, given {\color{black} its} willingness to pay $u_i$.

\begin{proposition}
\label{prop:buyersStrategies}
\textit{For given values of prices $(p_o,p_r)$ and quality/supply $(q_o,q_s)$, the payoff maximizing {\color{black}strategy for buyer $i$} will be:} 

\textit{a) if $q_s > q_o,$}
\begin{equation}
    b^*= \begin{cases} \text{sharing\ pool} & \text{if} \ u_i > \max(\frac{p_r-p_o}{q_s-q_o}, \frac{p_r}{q_s})\\
    \text{on-demand\ pool} & \text{if} \ \frac{p_o}{q_o}<u_i\leq \frac{p_o-p_r}{q_o-{\color{black}q_s}} \\
    \text{none} & \text{otherwise}
    \end{cases}
\end{equation}
\textit{b) if $q_o>q_s,$}
\begin{equation}
    b^*= \begin{cases} \text{sharing\ pool} & \text{if} \ \frac{p_r}{q_s}<u_i\leq \frac{p_r-p_o}{q_s-q_o} \\
    \text{on-demand\ pool} & \text{if} \ u_i > \max(\frac{p_o-p_r}{q_o-q_s},\frac{p_o}{q_o}) \\
    \text{none} & \text{otherwise}
    \end{cases}
\end{equation}
\textit{c) if $q_o=q_s,$}
\begin{equation}
    b^*= \begin{cases} \text{sharing\ pool} & \text{if} \ p_o >{\color{black}p_r}\ \&\ u_i >\frac{{\color{black}p_r}}{q_s} \\
    \text{on-demand\ pool} & \text{if} \ {\color{black}p_r} >p_o\ \&\ u_i > \frac{p_o}{q_o} \\
    \text{none} & \text{otherwise}
    \end{cases}
\end{equation}
\end{proposition}

{\color{black}\textit{Proof:} 
We compare $\pi_{i}^o(\mu_i)$, $\pi_{i}^s(\mu_i)$, and $0$, and find the conditions under which each strategy maximizes buyer $i$'s payoff.} When $\pi_{i}^o(\mu_i)$ is the largest, we have $b=\text{on-demand pool}$. When $\pi_{i}^s(\mu_i)$ is the largest, we have $b=\text{sharing pool}$. Otherwise, $b=\text{none}$.    \qed

\begin{figure*}[t]
  \centering
\subfigure[Both on-demand and sharing.]{\includegraphics[scale=0.37]{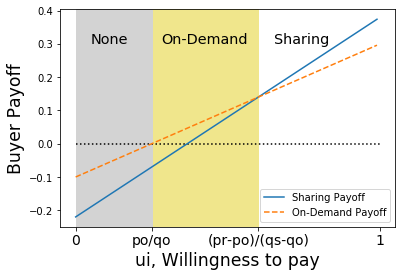}}
  \subfigure[Only On-demand.]
    {\includegraphics[scale=0.37]{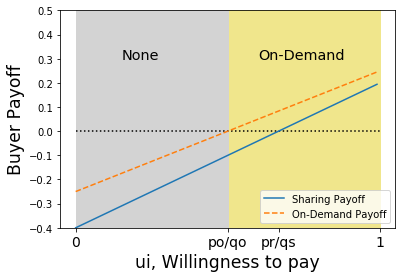}}
    \subfigure[Only Sharing.]
  {\includegraphics[scale=0.37]{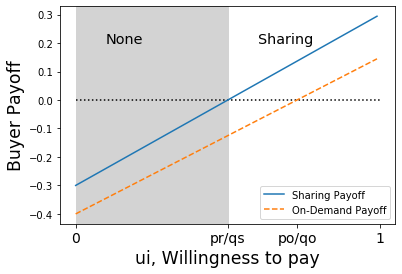}}
  \caption[users' Payoff at different willingness to pay, $q_s>q_o$.]
  {{\color{black}Illustrations of buyers' payoff, and therefore the decision they make,} at different willingness to pay, with $q_s>q_o$.
  }
  \label{fig:userPayoff}
\end{figure*}

\textcolor{black}{Note that Prop. \ref{prop:buyersStrategies} is independent of the distribution of $u_i$, users' willingness to pay.}
\textcolor{black}{Fig. \ref{fig:userPayoff}} provides a geometric description of Proposition \ref{prop:buyersStrategies} (Prop. \ref{prop:buyersStrategies}a in particular). For example, if $q_s>q_o$ (corresponding to Prop. \ref{prop:buyersStrategies}a), and when $u_i$ satisfies $\frac{p_o}{q_o} < u_i \leq \frac{p_r-p_o}{q_s-q_o}$, we can see from \textcolor{black}{Fig. \ref{fig:userPayoff}a} that the {\color{black}buyer's} payoff from the on-demand pool would be positive and higher than that from the sharing pool. {\color{black}Hence, the buyer} with $u_i$ satisfying $\frac{p_o}{q_o} < u_i \leq \frac{p_r-p_o}{q_s-q_o}$ would choose the on-demand pool.
Proposition \ref{prop:buyersStrategies} shows that buyers with higher willingness to pay $u_i$ will choose the option with higher quality/supply ($q_o$ or $q_s$ which is the gradient of all the payoff graphs in Fig. \ref{fig:userPayoff}), as we can see in all three graphs. For an example, {\color{black} we look at the case where $q_s>q_o$ in Fig. \ref{fig:userPayoff}. Users} with higher $u_i$ choose the sharing pool when $p_r-p_o \leq 0$ (Fig. \ref{fig:userPayoff}c), and also when {\color{black}$0 \leq \frac{p_r-p_o}{q_s-q_o} \leq 1$ (Fig. \ref{fig:userPayoff}a).} Nevertheless, we see that when the price {\color{black}$p_r$ for the sharing pool is too high such that $\frac{p_r-p_o}{q_s-q_o}>1$ (Fig. \ref{fig:userPayoff}b, where $\frac{p_r-p_o}{q_s-q_o}$ is the intersection of the two graphs),} none of the users will choose the sharing pool {\color{black}even with $q_s>q_0$}.

In the special case where $q_s=q_o$ (Proposition \ref{prop:buyersStrategies}c), the qualities/supply of the two options are the same. Geometrically, this means that the gradients of the two payoffs will be the same. If the on-demand pool has a higher price ($p_o>p_r$), the buyers will choose the sharing pool, and vice versa. 
When the buyers's willingness to pay is low ($u_i < \min(\frac{p_r}{q_s},\frac{p_o}{q_o})$), the buyer will choose neither.

\subsubsection{The MNO's price selection strategy}

Following the buyers' optimal selection strategies (Proposition \ref{prop:buyersStrategies}),
and integrating \textcolor{purple}{over} the collective behavior of all of the buyers, 
there will be 4 regions at equilibrium,
with boundary conditions as functions of $q_o$, $q_s$, $p_o$ and $p_r$. These 4 regions correspond to the 4 scenarios of whether \textcolor{red}{users will choose the sharing and/or on-demand option, or choose not to buy at all.} 
In Proposition \ref{prop:4regions}, we characterize these 4 regions.

\begin{proposition}
\label{prop:4regions}
For given values of prices ($p_r,p_o$) and quality/supply ($q_s,q_o$), the equilibrium will satisfy exactly one of the following 4 conditions: 
\begin{equation}
\begin{split}
    \text{R1} :\ \{(p_r,p_o) | p_r < \min(\frac{q_s}{q_o},1)p_o+(q_s-q_o)^+,\\ p_r > \max(\frac{q_s}{q_o},1)p_o-(q_o-q_s)^+ \}
    \label{eq:R1regioncond}
    \end{split}
\end{equation}
\begin{equation}
\label{eq:R2regioncond}
    \text{R2} :\ \{(p_r,p_o) | p_o < q_o,\ p_r \geq \min(\frac{q_s}{q_o},1) p_o +(q_s-q_o)^+ \}
\end{equation}
\begin{equation}
\label{eq:R3regioncond}
    \text{R3} :\ \{(p_r,p_o) | p_r < q_s,\ p_r \leq \max(\frac{q_s}{q_o},1)p_o-(q_o-q_s)^+ \} 
\end{equation}
\begin{equation}
\label{eq:R4regioncond}
    \text{R4} :\ \{(p_r,p_o) | p_r\geq q_s,\ p_o\geq q_o \},
\end{equation}
where $(a)^+=a$ if $a\geq 0$, and $0$ otherwise. 
In R1, buyers buy from both pools {\color{black} or neither, like in Fig. \ref{fig:userPayoff}a}. 
In R2, buyers buy from the on-demand pool {\color{black}or from neither, as in Fig. \ref{fig:userPayoff}b}.
In R3, buyers buy from the sharing pool {\color{black}or from neither, as in Fig. \ref{fig:userPayoff}c}.
In R4, buyers will buy from neither pool. 
\end{proposition}

{\color{black}\textit{Proof:} 
The boundary conditions of the four regions are deduced through Fig. \ref{fig:userPayoff}. \qed}

\textcolor{black}{Prop. \ref{prop:4regions} is independent of the distribution of $u_i$, users' willingness to pay.}
This {\color{black} proposition} indicates how the relative magnitudes of the prices $p_r$, $p_o$, and supply $q_s$ and $q_o$ impact the equilibrium outcome. 
Specifically, when $(p_r,p_o,q_s,q_o)$ satisfies the boundary conditions of one of the regions, the collective behavior of buyers (who behave according to {\color{black}Proposition \ref{prop:buyersStrategies}}), will result in the equilibrium outcome being in that region.  
For example, when the prices $p_o$ and $p_r$ satisfy the conditions $p_o < q_o,\ p_r \geq \min(\frac{q_s}{q_o},1) p_o +(q_s-q_o)^+$, 
it will not be incentive compatible for any buyers to choose the sharing pool (regardless of $u_i$), and the outcome is R2. 
The special case when $q_s=q_o$ in Proposition \ref{prop:buyersStrategies} also fits this 4 region framework. If $1 < \min(\frac{p_r}{q_s},\frac{p_o}{q_o})$, the users will choose not to buy and the resulting scenario will be R4.
In the alternative where $1 < \min(\frac{p_r}{q_s},\frac{p_o}{q_o})$ does not hold, if the on-demand price is higher ($p_o>p_r$) the users will choose the sharing pool and we end up in R3, and vice versa for R2.
\textcolor{violet}{This 4 region structure is unique with respect to the results from work on other sharing platforms in networked systems (mobile data re-selling \cite{mobileDataTrading1,mobileDataTrading3} and crowdsourced wireless community networks \cite{wifiSharing}).}


With the 4 disjoint regions involving different pools being purchased from (Proposition \ref{prop:4regions}), the revenue function $R(p_o,p_r)$ can be simplified as follows: 
\begin{equation}
\begin{split}
    & R^{R1}(p_o,p_r)=\\ & \begin{cases}
    & N p_o(\frac{p_r-p_o}{q_s-q_o}-\frac{p_o}{q_o})+ N p_r \delta (1-\frac{p_r-p_o}{q_s-q_o}), \quad \text{if} \ q_s\geq q_o\\
    & N p_r \delta (\frac{p
    _o-p_r}{q_o-q_s}-\frac{p_r}{q_s})+ N p_o (1-\frac{p
    _o-p_r}{q_o-q_s}),\quad  \text{if} \ q_o >q_s 
    \end{cases}
    \end{split}
\end{equation}
\begin{equation}
    R^{R2}(p_o,p_r)=N p_o(1-\frac{p_o}{q_o})
\end{equation}
\begin{equation}
    R^{R3}(p_o,p_r)=N p_r \delta (1-\frac{p_r}{q_s})
\end{equation}
\begin{equation}
    R^{R4}(p_o,p_r)=0
\end{equation}
In R1, buyers are willing to buy from both the on-demand and sharing pool. The two terms in the objective $R^{R1}(p_o,p_r)$ correspond to the revenue obtained from the two pools. In R1 with $q_s\geq q_0$, $(\frac{p_r-p_o}{q_s-q_o}-\frac{p_o}{q_o})$ and $1-\frac{p_r-p_o}{q_s-q_0}$ are the proportion of the buyers who choose the on-demand and sharing pool respectively. In R4, buyers find it optimal to not purchase from either pool, and hence the MNO's revenue is $0$.

Therefore, to optimize the revenue maximization problem, we will split the problem into four subproblems as follows:
\begin{equation}
    \underset{p_o,p_r}{\text{max}}\ {\color{black}R^{R1}(p_o,p_r)}\ \text{s.t.}\ (\ref{eq:R1regioncond})
\end{equation}
\begin{equation}
    \underset{p_o,p_r}{\text{max}}\ {\color{black}R^{R2}(p_0,p_r)}\ \text{s.t.}\ (\ref{eq:R2regioncond})
\end{equation}
\begin{equation}
    \underset{p_o,p_r}{\text{max}}\ {\color{black}R^{R3}(p_0,p_r)}\ \text{s.t.}\ (\ref{eq:R3regioncond})
\end{equation}
\begin{equation}
    \underset{p_o,p_r}{\text{max}}\ {\color{black}R^{R4}(p_0,p_r)\ \text{s.t.}\ (\ref{eq:R4regioncond})}
\end{equation}

For subproblem $R^{R1}(p_o,p_r)$, we further split it into 2 subproblems $R^{R1a}(p_o,p_r)$ and $R^{R1b}(p_o,p_r)$ corresponding to the 2 cases where {\color{black}$q_s\geq q_o$} and $q_o> q_s$. 
{\color{black}In the following, we give Theorem 1 to characterise} the optimal prices $p_o$ and $p_r$.

\begin{theorem}
\label{thr:buyerGameOpt}
When buyers make decisions according to Proposition \ref{prop:buyersStrategies}, given a fixed $(q_o,q_s)$, and if the condition $(\delta+1)^2 q_s <4 \delta q_o$ holds, the MNO's optimal prices will be as follows.

\begin{itemize}
\item To maximize $R^{R1a}(p_0,p_r)$ in R1, we have:

If Eq. (\ref{eq:R1regioncond}) is satisfied 
\begin{equation*}
\begin{split}
& (p_{o,1a}^*,p_{r,1a}^*)= \Bigg(\frac{\delta (1+\delta)q_o(q_s-q_o)}{4q_s\delta-(1+\delta)^2q_o},\\
& \qquad\qquad\qquad\quad \frac{q_s-q_o}{2}+\frac{(1+\delta)^2q_o(q_s-q_o)}{2(4q_s\delta -(1+\delta)^2q_o)}\Bigg),
\end{split}
\end{equation*}
Else,
\begin{equation*}
(p_{o,1a}^*,p_{r,1a}^*)\!=\! \text{argmax}_{\{(\frac{q_o}{2}, q_s-\frac{q_o}{2}),(\frac{q_o}{2},\frac{q_s}{2}) \}} R^{R1a}(p_o,p_r). 
\end{equation*}
    
\item To maximize $R^{R1b}(p_o,p_r)$ in R1, we have:

If {\color{black}Eq. (\ref{eq:R1regioncond})} is satisfied,
\begin{equation*}
    (p_{o,1b}^*,p_{r,1b}^*)\!=\! \!(\frac{2\delta q_o(q_o-q_s)}{4\delta q_o-(\delta+1)^2q_s}, \frac{q_s(\delta+1)(q_o-q_s)}{4\delta q_o-(\delta+1)^2q_s}),
\end{equation*}
Else,
\begin{equation*}
   (p_{o,1b}^*,p_{r,1b}^*)= \text{argmax}_{\{ (q_o-\frac{q_s}{2}, \frac{q_s}{2}),(\frac{q_o}{2},\frac{q_s}{2}) \}} R^{R1b}(p_o,p_r).
\end{equation*}

\item To maximize $R^{R2}(p_o,p_r)$, we have $(p_{o,2}^*,p_{r,2}^*)=(\frac{q_o}{2}, a)$.\\
 
\item To maximize $R^{R3}(p_o,p_r)$, we have $(p_{o,3}^*,p_{r,3}^*)=(b, \frac{q_s}{2})$.
\end{itemize}
{\color{black}Note that, in the latter two cases,} $a$ is any price $p_r$ which satisfies the boundary conditions of R2 (Eq. (\ref{eq:R2regioncond})), 
and $b$ is any price $p_o$ which satisfies the boundary conditions of R3 (Eq. (\ref{eq:R3regioncond})).
\end{theorem}

{\color{black}\textit{Proof:} See Appendix 
For each of the subproblems, 
we obtain the Hessian matrix of the objective function. The Hessians of {\color{black}R1a, R2 and R3} are all negative semidefinite (having eigenvalues which $\leq 0$), and therefore these {\color{black}subproblems} have concave objectives. As the constraints are linear with respect to $p_o$ and $p_r$, these {\color{black}subproblems} are convex. For the {\color{black}subproblem} {\color{black}R1b}, the Hessian is negative semidefinite under the condition $(\delta+1)^2 q_s <4 \delta q_o$.
With convexity, we can use the KKT conditions {\color{black}to find their optimal prices}.
\qed}

{\color{black}In Theorem 1,} $(\delta+1)^2 q_s <4 \delta q_o$ is the condition under which the Hessian of $R1b$'s objective has negative eigenvalues, and hence the condition under which subproblem $R1b$ is convex.
We use the KKT conditions to solve the problem. For subproblem $R1a$, the gradients are $0$ at $p^*_{o,1a}=\frac{2\delta q_o(q_o-q_s)}{4\delta q_o-(\delta+1)^2q_s}$ and $p^*_{r,1a}=\frac{q_s(\delta+1)(q_o-q_s)}{4\delta q_o-(\delta+1)^2q_s}$. If this point is within the boundary conditions, it is the optimal point. If it is not, the optimal point would be at one of the boundaries (with regions $R2$ or $R3$), specifically being $(\frac{q_o}{2}, q_s-\frac{q_o}{2})$ (boundary with $R2$) or $(\frac{q_o}{2},\frac{q_s}{2})$ (boundary with $R3$). Likewise for subproblem $R1b$.
For subproblems $R2$ and $R3$, the optimal solutions of $p_{o,2}^*=\frac{q_o}{2}$ and $p_{r,3}^*=\frac{q_s}{2}$ hold at the boundary as well.
These optimal prices agree with intuition, that as the 
{\color{black}quality/supply} (likelihood of buyer getting service) $q_o$ (or $q_s$) increases, the MNO can increase the price to increase its revenue.

\subsection{Case 2: Re-sellers weigh inconvenience cost when deciding to re-sell quota}
\label{subsec:combined}

In this subsection, we analyse \textcolor{black}{an alternative platform design in which the MNO implements manual decision making for re-sellers:} re-sellers weigh their inconvenience cost $g_i$ when deciding whether or not to re-sell quota in the sharing pool, according to Eq. (\ref{sellersPayoffEqns}). 
As mentioned in Section \ref{sec:model}, the inconvenience cost $g_i$ factors in the re-seller's uncertainty of usage during the timeslot, and the re-seller's inconvenience, willingness and ability towards checking the app to decide whether or not to re-sell.
Here $g_i$ follows the uniform $U[0,1]$ distribution as it captures the varying levels of inconvenience across users. In Section \ref{sec:simulations}, we perform simulations using both uniform and beta (which has a gaussian-like structure) distributions, showing similar results across distributions.

Under this scenario, we solve the Stackelberg Game (Eq. (\ref{LeaderEq}), (\ref{Follower1Eq}) and (\ref{Follower2Eq})).
Firstly, in Stage 1 the MNO decides the prices $p_r$ and $p_o$. Following which, in Stage $2A$ the re-sellers decide whether they want to sell their excess resource quota, based on what maximizes their payoff (Eq. (\ref{sellersPayoffEqns})), resulting in a {\color{black}quality/supply} of $q_s(p_r)$.
Finally, in Stage $2B$, the buyers would make a decision, aiming to maximize their individual payoffs (Eq. (\ref{eq:BuyerPayoff})), given the values of $(p_r,p_o,q_s(p_r),q_o)$.
We solve the Stackelberg Game by backward induction, solving Stage $2$ first, and using the equilibrium outcome of Stage $2$ to solve Stage $1$ and obtain the MNO's optimal prices. 

\subsubsection{The re-seller's optimal strategy}

The following Proposition shows the re-seller's optimal strategy.
\begin{proposition}
\label{SellersOptimalStrategy}
For given $p_r$, {\color{black}re-seller $i$} with inconvenience cost $g_i$'s optimal strategy would be:
\begin{equation}
    s^*= \begin{cases}
    & sell, \quad \text{if}\ (1-\delta)p_r>g_i,\\
    & no,\quad \text{otherwise}
    \end{cases}
\end{equation}
\end{proposition}
Collectively, by Eq. (\ref{propEqn}) this results in a proportion $(1-\delta)p_r$ of re-sellers being willing to rent out their unused resources and therefore {\color{black}following Eq. (\ref{eq:supplyGeneric}), the total quality/supply of} the sharing pool will be
\begin{equation}
    q_s(p_r)\sim f( (1-\delta)p_r ),
\end{equation}
{\color{black}which is a function of price $p_r$.}



\subsubsection{The buyer's optimal strategy}
Given $(q_s(p_r),q_o)$ and $(p_r,p_o)$, a buyer's strategy would still follow Proposition \ref{prop:buyersStrategies} in {\color{black}Section} \ref{subsec:MNObuyers}, with the substitution of $q_s$ by $q_s(p_r)$. 
Hence, their collective strategies would form distinct regions R1 to R4 at equilibrium, where the four regions indicate whether or not the sharing and/or on-demand pools are feasible. This is according to Proposition \ref{prop:4regions} in {\color{black}Section} \ref{subsec:MNObuyers}. Once again, the only difference is that $q_s$ is replaced by $q_s(p_r)$. 

\subsubsection{The MNO's optimal price}
Given the strategies of the re-sellers and buyers in Stages $2A$ and $2B$, corresponding to the regions in Proposition
\ref{prop:4regions}, the MNO's revenue maximizing problem $R(p_o,p_r)$ can be split into the following subproblems. 

$R^{R1a}_{\text{C2}}(p_o,p_r)$:
\begin{equation}
\begin{aligned}
    \underset{p_r,p_o}{\text{max}}\ & N p_o(\frac{p_r-p_o}{q_s(p_r)-q_o}-\frac{p_o}{q_o})+ N p_r \delta (1-\frac{p_r-p_o}{q_s(p_r)-q_o})\\
    \text{s.t.}\ &  (\ref{eq:R1regioncond}),\;
    q_s(p_r) > q_o 
    \end{aligned}
\end{equation}
$R^{R1b}_{\text{C2}}(p_o,p_r)$:
\begin{equation}
    \begin{aligned}
    \underset{p_r,p_o}{\text{max}}\ & N p_r \delta (\frac{p
    _o-p_r}{q_o-q_s(p_r)}-\frac{p_r}{q_s(p_r)})
    + N p_o (1-\frac{p
    _o-p_r}{q_o-q_s(p_r)})\\ 
    \text{s.t.}\ &  (\ref{eq:R1regioncond}),\; 
    q_o > q_s(p_r)
    \end{aligned}
\end{equation}
\begin{equation}
    \begin{aligned}
    R^{R2a}_{\text{C2}}(p_o,p_r):\quad \underset{p_r,p_o}{\text{max}}\ & N p_o(1-\frac{p_o}{q_o})\\
    \text{s.t.}\ & (\ref{eq:R2regioncond}),\; q_s(p_r) > q_o,
    \end{aligned}
\end{equation}
\begin{equation}
    \begin{aligned}
    R^{R2b}_{\text{C2}}(p_o,p_r):\quad\underset{p_r,p_o}{\text{max}}\ & N p_o(1-\frac{p_o}{q_o})\\
    \text{s.t.}\ & (\ref{eq:R2regioncond}),\; q_o > q_s(p_r),\\ 
    \end{aligned}
\end{equation}
\begin{equation}
    \begin{aligned}
    R^{R3a}_{\text{C2}}(p_o,p_r):\quad \underset{p_r,p_o}{\text{max}}\ & N p_r \delta (1-\frac{p_r}{q_s(p_r)})\\
    \text{s.t.}\ & (\ref{eq:R3regioncond}),\: q_s(p_r) > q_o,\\ 
    \end{aligned}
\end{equation}
\begin{equation}
    \begin{aligned}
    R^{R3b}_{\text{C2}}(p_o,p_r):\quad \underset{p_r,p_o}{\text{max}}\ & N p_r \delta (1-\frac{p_r}{q_s(p_r)})\\
    \text{s.t.}\ & (\ref{eq:R3regioncond}),\; q_o > q_s(p_r), 
    \end{aligned}
\end{equation}
and $R^{R4}_{\text{C2}}(p_o,p_r)$ is a constant of $0$, corresponding to R4 in Proposition \ref{prop:4regions} (where users choose neither the on-demand nor sharing pool). Other than {\color{black}R4}, the other regions are considered {\color{black}with} two subproblems which correspond to the disjoint cases {\color{black}$q_s(p_r)\geq q_o$} and $q_o>q_s(p_r)$. 
This allows us to replace the terms involving $min()$ and $()^+$.
For example, {\color{black}$R^{R1a}_{\text{C2}}(p_o,p_r)$ and $R^{R1b}_{\text{C2}}(p_o,p_r)$} both correspond to the situation when we are in R1 (where at equilibrium, buyers choose both to buy from the on-demand pool and the sharing pool). They differ depending on whether $q_s(p_r) \geq q_o$ {\color{black}or not}. 


The subproblems are not jointly convex with respect to $p_o$ and $p_r$. {\color{black}Hence, with $p_r$ and $q_s(p_r)$ given}, we optimize with respect to $p_o$ to attain insights on the MNO's optimal price {\color{black}in R1 and R2. The MNO's revenues in R3 and R4 are not related to $p_o$ \textcolor{purple}{(since no users buy from the on-demand pool)} and are constant with $p_r$ and $q_s(p_r)$ given.}
\begin{theorem} \label{thr:optimalPo}
Given {\color{black}$(p_r,q_s(p_r))$}, the MNO's optimal on-demand price $p_o^*$ at each region is given by:
\begin{itemize}
    \item In R1 with $q_s(p_r)\geq q_o$,
        \begin{equation*}
            p_{o,1a}^*=\frac{q_o p_r (1+\delta)}{2 q_s(p_r,\delta)} \text{ or } \text{argmax}_{\{ B_1,B_2 \}} {\color{black}R^{R1a}_{\text{C2}}(p_o,p_r)}
        \end{equation*}
    \item In R1 with $q_s(p_r)<q_o$,
        \begin{equation*}
        \begin{split}
            & p_{o,1b}^*=\frac{p_r(\delta+1)+q_o-q_s(p_r,\delta)}{2}\\
            & \text{ or }\text{argmax}_{\{B_1,B_2\}} {\color{black}R^{R1b}_{\text{C2}}(p_o,p_r)}
        \end{split}
        \end{equation*}
    \item In R2 with $q_s(p_r)\geq q_o$,\ $p_{o,2a}^*= \frac{q_o}{2} \text{ or } B_1$\\
        
    \item In R2 with $q_s(p_r)<q_o$,\ $p_{o,2b}^*= \frac{q_o}{2} \text{ or } B_1$
\end{itemize}
where $B_1=p_r-q_s(p_r)+q_o$ and $B_2=\frac{p_r q_o}{q_s(p_r)}$ are points on the region boundaries.
\end{theorem}

\textit{Proof:}
For subproblem {\color{black}$R^{R1a}_{\text{C2}}$}, the second derivative of the objective function with respect to $p_o$ is $\frac{-2N q_s(p_r)}{(q_s(p_r)-q_o)q_o} <0$ since {\color{black}$q_s(p_r)\geq q_o$} and $q_o>0$, indicating that the objective function is concave. As the constraints involve linear functions of $p_o$, the maximization problem is convex with respect to $p_o$.
Therefore, we convert the constraints to non-strict inequalities and use the KKT conditions to solve {\color{black}$R^{R1a}_{\text{C2}}$}, with respect to $p_o$. The Lagrangian is defined as $L(p_o,\lambda_1, \lambda_2)=N p_o(\frac{p_r-p_o}{q_s(p_r)-q_o}-\frac{p_o}{q_o})+N p_r \delta (1-\frac{p_r-p_o}{q_s(p_r)-q_o}) +\lambda_1(p_r-p_o-q_s(p_r)+q_o) + \lambda_2(q_a(p_r)/q_o p_o-p_r)$, with $\lambda_1$ and $\lambda_2$ being the dual variables corresponding to the constraints. If the dual variables $\lambda_1$ (or $\lambda_2$) were greater than $0$, the optimal point would lie on the boundary between region {\color{black}R1 and R2 (or R3 respectively)} 
If the dual variables are $0$, 
the optimal point will be the point at which the gradient of the Lagrangian is $0$.
We solve the rest of the subproblems similarly.
\qed

Theorem \ref{thr:optimalPo} gives us insights on how the {\color{black}MNO} should set the on-demand price $p_o$, given that the other parameters {\color{black}$p_r$ and $q_s(p_r)$} are fixed. In R1, when users buy from both the sharing and on-demand pool,
the price $p_o$ at which the gradient of the revenue is $0$ is $\frac{q_o p_r (1+\delta)}{2 q_s(p_r)}$ for R1a and $\frac{p_r(\delta+1)+q_o-q_s(p_r)}{2}$ for R1b. 
\textcolor{red}{We seperately analyse the cases when $\frac{q_o p_r (1+\delta)}{2 q_s(p_r)}$ (the point at which the gradient of the revenue is $0$) lies within the region's boundaries, and when it does not.}
\underline{Case 1:} If $p_o^*=\frac{q_o p_r (1+\delta)}{2 q_s(p_r)}$ lies within R1a's boundaries ($p_r < p_o +q_s(p_r)-q_o, p_r > \frac{q_s(p_r)}{q_o}p_o$), it is the optimal price, as the problem is convex with respect to $p_o$. Likewise for $\frac{p_r(\delta+1)+q_o-q_s(p_r)}{2}$ in R1b.
These functions indicate that the optimal price $p_o^*$ increases with respect to the on-demand expected quality (supply) $q_o$ and platform commission $\delta$. As the on-demand supply $q_o$ increases, the increase in on-demand price $p_o$ would not deter users away from the on-demand pool. Likewise, when the MNO's commission $\delta$ increases, re-sellers will be deterred from selling their unused resource quota, resulting in a smaller supply at the sharing-pool. With a potentially lower payoff at the sharing pool, the increase in the on-demand price $p_o$ would not deter users away from the on-demand pool.
At the same time, as the sharing supply {\color{black}$q_s(p_r)$} increases, the sharing pool is seen as more attractive to users (due to the potentially higher individual payoff). In light of this, the optimal on-demand price $p_o^*$ has to decrease. 
\underline{Case 2:} If $\frac{q_o p_r (1+\delta)}{2 q_s(p_r)}$ (and $\frac{p_r(\delta+1)+q_o-q_s(p_r)}{2}$) do not lie within R1a's (respectively R1b's) boundaries, the optimal points $p_{o,1a}^*$ (and $p_{o,1b}^*$) will lie on the region's boundaries, since the problems are convex. Specifically, the optimal prices will be {\color{black}$\text{argmax}_{\{B_1,B_2 \}} R^{R1a}_{\text{C2}}(p_o,p_r)$ (and $\text{argmax}_{\{B_1,B_2 \}} R^{R1b}_{\text{C2}}(p_o,p_r)$)}.

In {\color{black}R2}, where {\color{black}buyers} only choose the on-demand pool, the price $p_o$ at which the gradient of the revenue is $0$ is $\frac{q_o}{2}$.
\textcolor{red}{We seperately analyse the cases when $\frac{q_o}{2}$ (the point at which the gradient of the revenue is $0$) lies within the region's boundaries, and when it does not.}
\underline{Case 1:} If $p_o^*=\frac{q_o}{2}$ lies within {\color{black}R2}'s boundary ($p_r \geq  p_o +(q_s(p_r)-q_o)$ for {\color{black}$R^{R2a}_{\text{C2}}(p_o,p_r)$} and  $p_r \geq \frac{q_s}{q_o} p_o$ {\color{black}for $R^{R2b}_{\text{C2}}(p_o,p_r)$} respectively), it is the optimal on-demand price because the problem is convex with respect to $p_o$. Therefore, the optimal on-demand price is an increasing function of the on-demand pool's supply $q_o$. This indicates that as the supply of the on-demand pool $q_o$ increases, the rise in the on-demand price will not deter users away from the on-demand pool. \underline{Case 2:} If $p_o^*= \frac{q_o}{2}$ does not lie within region {\color{black}R2}'s boundaries, the optimal point will lie on the boundary $p_r = B_1$ for $R^{R2a}_{\text{C2}}(p_o,p_r)$ and $p_r = B_2$ for $R^{R2b}_{\text{C2}}(p_o,p_r)$, respectively. {\color{black}This is because, the optimization problems are convex with respect to $p_o$.} 

{\color{black}On this basis, we then focus on the joint optimization of $p_o$ and $p_r$ for the MNO's optimal revenue, which will be the maximum of the revenues across the four regions. According to Corollary 1, the MNO's optimal revenue is calculated as} 
\begin{equation}
    \max_{p_r} \quad R_{\max}(p_r).
    \label{eq:optrevenue}
\end{equation}

\begin{corollary}
\label{corollary:OptRev}
Given {\color{black}$(p_r,q_s(p_r))$}, the MNO's optimal revenue will be
\begin{equation}
\begin{split}
    & {\color{black}R_{\max}(p_r)}=\max\Bigg\{ R^{R1a}_{\text{C2}}(p_{o,1a}^*,p_r), R^{R1b}_{\text{C2}}(p_{o,1b}^*,p_r),\\ & \qquad \qquad \qquad R^{R2a}_{\text{C2}}(p_{o,2a}^*,p_r), R^{R2b}_{\text{C2}}(p_{o,2b}^*,p_r), R^{R3}_{\text{C2}}(p_r),0\Bigg\},
    \end{split}
\end{equation}
where $R^{R3}_{\text{C2}}(p_r)=N p_r \delta (1-\frac{p_r}{q_s(p_r)})$ is the revenue in R3 (the region where the buyers only choose the sharing pool), if the boundary conditions of R3 (Eq. (\ref{eq:R3regioncond})) were met. And $0$ is the revenue in R4 (the region where the buyers choose neither the sharing pool nor the on-demand pool), if the boundary conditions of R4 (Eq. (\ref{eq:R4regioncond})) were met.
\end{corollary}

{\color{black}Since the optimal on-demand price $p_o^*$ in $R_{\max}(p_r)$ is obtained in Theorem 2, the optimization in (\ref{eq:optrevenue}) can be addressed} by using a linear search across $p_r$. In Section \ref{sec:simulations}, we discuss the joint optimization of $p_o$ and $p_r$ and the resulting insights at the optimal and equilibrium point.

\section{Numerical Simulations} \label{sec:simulations}
\begin{figure*}[t]
  \centering
\subfigure[The MNO's optimal revenue and region against $q_o$. $\delta=0.2, a=2$, with $u_i,g_i$ following the uniform $U(0,1)$ distribution. ]{\includegraphics[scale=0.37]{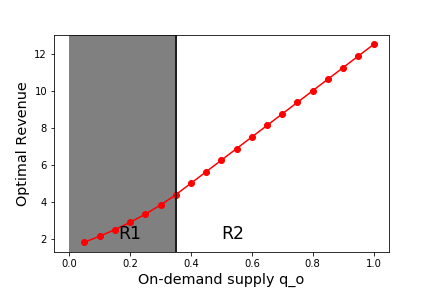}}
\subfigure[The MNO's optimal revenue and region against $q_o$. $\delta=0.2, a=2$, with $u_i,g_i$ following \textcolor{black}{beta(2,2)} distribution.]{\includegraphics[scale=0.37]{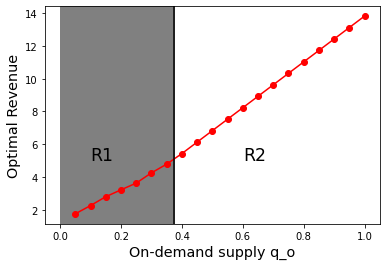}}
\subfigure[\textcolor{blue}{The MNO's optimal revenue and region against $q_o$. $\delta=0.2, a=2$, with $u_i,g_i,$ and usage levels following the uniform $U(0,1)$ distribution.}]{\includegraphics[scale=0.37]{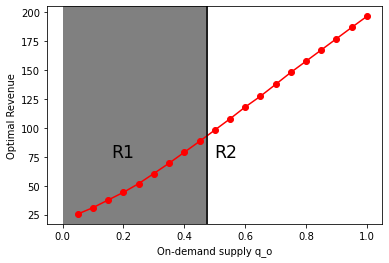}}
\caption[Numerical Simulations and Insights]
  {\textcolor{blue}{When on-demand supply is low, the optimal revenue is at Region 1 (offering both a sharing and an on-demand pool). As the on-demand supply increases, the optimal revenue transitions to Region 2 (purely on-demand).}} 
  \label{fig:impactQo}
\end{figure*}
\begin{figure*}
\centering
  \subfigure[The MNO's optimal revenue against on-demand supply $q_o$, $a=2$.]
   {\includegraphics[scale=0.37]{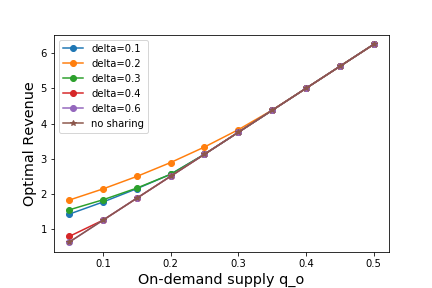}}
    \subfigure[The MNO's optimal revenue against commission $\delta$, $a=2$.]
  {\includegraphics[scale=0.37]{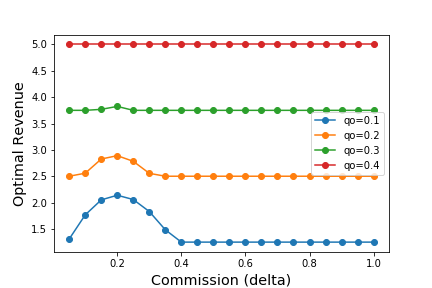}}
  \subfigure[The MNO's optimal revenue and region against $\delta$. $q_o=0.2, a=2$, with $u_i, g_i$ following the uniform $U(0,1)$ distribution.]
  {\includegraphics[scale=0.37]{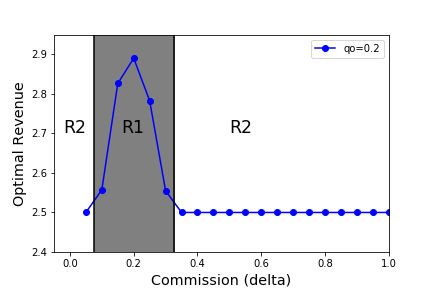}}
  \subfigure[The MNO's optimal revenue and region against $\delta$. $q_o=0.2, a=2$, with $u_i, g_i$ following the \textcolor{black}{beta(2,2)} distribution.]
  {\includegraphics[scale=0.37]{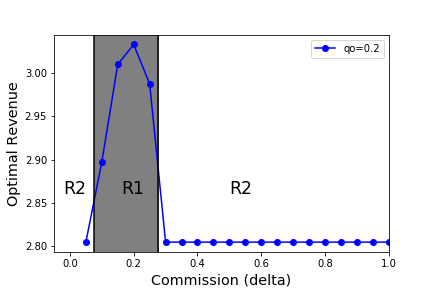}}
  \subfigure[\textcolor{blue}{The MNO's optimal revenue and region against $\delta$. $q_o=0.2, a=2$, with $u_i, g_i$, and usage levels following the uniform $(0,1)$ distribution.}]
  {\includegraphics[scale=0.37]{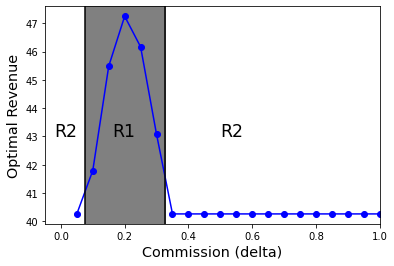}}
  \caption[Numerical Simulations and Insights]
  {\textcolor{blue}{When the commission level $\delta$ is lower, the optimal revenue is at Region 1 (offering both a sharing and an on-demand pool). As the commission level increases, the optimal revenue transitions to Region 2 (purely on-demand).}  }
  \label{fig:impactDelta}
\end{figure*}

\begin{figure*}
\subfigure[\textcolor{black}{The MNO's revenue and region against on-demand price $p_o$. $q_o, \delta=0.2, a=2, p_r=0.5$, with $u_i, g_i$ following the beta(2,2) distribution.}]
 {\includegraphics[scale=0.37]{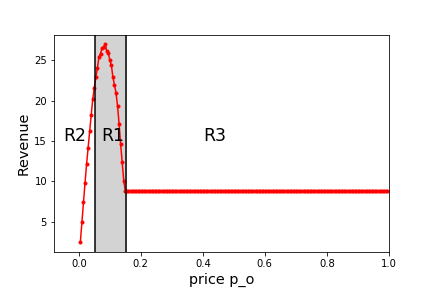}}
  \subfigure[The MNO's optimal revenue and region against on-demand supply $q_o$, $\delta=0.2, a=1.5 $.]
  {\includegraphics[scale=0.37]{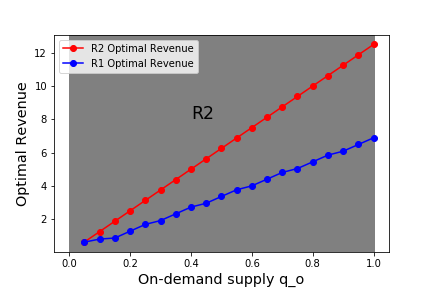}}
  \subfigure[The individual buyer's payoff against willingness to pay $u_i$, $\delta=0.2,a=2$.]{\includegraphics[scale=0.37]{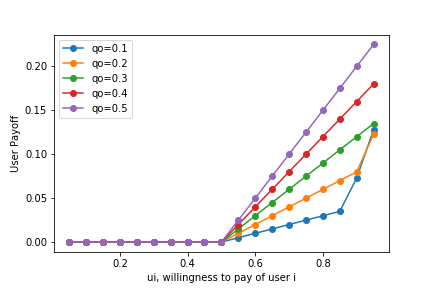}}

  \caption[Numerical Simulations and Insights]
  {Further Numerical Results and Insights: The impact of the on-demand price and normalization parameter on revenue, and how the individual buyer's payoff varies with the willingness to pay. } 
 
  \label{fig:combinedSimul}
\end{figure*}

In this section we provide numerical results on the optimal and equilibrium outcome of the Stackelberg Game (Eq. (\ref{LeaderEq}), (\ref{Follower1Eq}) and (\ref{Follower2Eq})). These results give us insights and show the practical implications of integrating the sharing pool with the On-Demand Pricing and Resource Reservation model.
\textcolor{red}{In many of our experiments, for each combination of $($on-demand supply $q_o,$ commission $\delta,$ and normalization parameter $a)$, we optimize over the prices $(p_o, p_r)$, obtaining the optimal revenue, and hence optimal region.} 
\textcolor{violet}{In Section \ref{sec:prototype}, we perform preliminary user studies on a prototype, to evaluate the extent of which user decisions differ from our decision model. }

Firstly, we define a specific form of $q_s(p_r)$, which indicates the relationship between the price $p_r$ and the resulting {\color{black}quality/supply $q_s(p_r)$, as follows:} 

\begin{equation}
    q_s(p_r)=a \log(1+(1-\delta)p_r)
    \label{eq:specificformQs}.
\end{equation}
We use a log function, which models the law of diminishing marginal returns \cite{shakkottai2008network}. 
The relative supply/quality (i.e. likelihood of buyers getting their jobs served) of the sharing pool increases as the proportion of re-sellers willing to sell $(1-\delta)p_r$ increases. 
The parameter $a$ is the \textcolor{brown}{normalization parameter, normalizing the function of the proportion of re-sellers $log(1+(1-\delta)p_r)$ willing to re-sell to a magnitude such that the supply level $q_s(p_r)$ is comparable with the supply at the on-demand pool $q_o$.} 
For our numerical simulations, we use Eq. (\ref{eq:specificformQs}) as the specific form of $q_s(p_r)$ and set the number of buyers and the number of re-sellers ($N$) to be  \textcolor{black}{50}. \textcolor{black}{We perform experiments with the willingness to pay $u_i$ and the willingness to share $g_i$ following the Uniform[0,1] as well as the Beta(2,2) distribution. Under the Uniform[0,1] distribution, the users' types $u_i$ and $g_i$ are evenly spread out over the range $[0,1]$. Under the Beta(2,2) distribution, the users' types follow an inverse parabolic structure, with more users whose types take values towards the center of the range $[0,1]$ and fewer users taking values towards the edge of the range $[0,1]$, somewhat similar to a Gaussian distribution.}
\textcolor{blue}{We also model heterogeneous user usage levels, both on the buyer side and the re-seller side, with usage levels following the uniform[0,1] distribution.}

\textcolor{red}{\textit{Impact of the on-demand supply $q_o$ on the optimal revenue and region :}}
Firstly, we show how the MNO's optimal revenue {\color{black}and region vary} with respect to the on-demand quality/supply $q_o$ in Fig. \ref{fig:impactQo}a, \ref{fig:impactQo}b and \ref{fig:impactQo}c
where we set $\delta=0.2$ and $a=2$. \textcolor{brown}{Given that we do not know what is an appropriate normalization parameter which relates the supply function $q_s(p_r)$ to the supply of the on-demand pool $q_o$, we set $a=2$ first, and later decrease the normalization parameter to $a=1.5$.} 
\textcolor{red}{To obtain the optimal revenue and region, we optimize over the prices $(p_o, p_r)$, given each combination of $($on-demand supply $q_o,$ commission $\delta,$ and normalization parameter $a)$.}
\textcolor{black}{For both cases where the re-sellers' willingness to share and the buyers' willingness to pay follow uniform (Fig. \ref{fig:impactQo}a) and beta distribution (Fig. \ref{fig:impactQo}b),} \textcolor{blue}{and when the users' usage levels are heterogeneous and follow the Uniform[0,1] distribution (Fig. \ref{fig:impactQo}c),} it can be seen that for smaller values of $q_o$, i.e., when the on-demand supply is lower, it is more beneficial to supplement buyers with the sharing pool to buy quota for job computation {\color{black}(Region R1)}. This additional income adds to the revenue of the MNO.
Nevertheless, as $q_o$ increases, the additional income from the sharing pool does not outweigh the increase in revenue from the on-demand pool $N_{\text{on-demand }} p_o (1-\frac{p_o}{q_o})$, hence {\color{black}Region R2} is optimal.

\textcolor{red}{\textit{Impact of the commission $\delta$ on the optimal revenue and region:}}
In Fig. \ref{fig:impactDelta}a, we plot the optimal revenue for the MNO against $q_o$, for different values of commission $\delta$. Conversely, in Fig. \ref{fig:impactDelta}b, we plot the optimal revenue of the MNO against the commission $\delta$, for different values of $q_o$. 
In Figs. \ref{fig:impactDelta}c, d and e, we show the optimal region at equilibrium for different values of $\delta$, \textcolor{black}{(with $g_i$ and $u_i$ taking values from the Uniform[0,1] and Beta(2,2) distributions,  \textcolor{blue}{and when the users' usage levels are heterogeneous following Uniform[0,1]} respectively).}
\textcolor{red}{Once again, to obtain the optimal revenue and region, we optimize over the prices $(p_o, p_r)$, given each combination of $($on-demand supply $q_o,$ commission $\delta,$ and normalization parameter $a)$.}
These graphs show that when the commission $\delta$ is smallest, it is not profitable to have sharing (hence the optimal region is R2), because the platform does not get much revenue from sharing due to its commission being small.
As $\delta$ gets larger, it becomes profitable to have the sharing platform (Region R1) due to the increase in commission for the platform and hence revenue $N_{\text{on-demand}} p_o(\frac{p_r-p_o}{q_s(p_r)-q_o}-\frac{p_o}{q_o})+N_{\text{share}} p_r \delta (1-\frac{p_r-p_o}{q_s(p_r)-q_o})$. On the other hand, for larger $\delta$, the re-sellers will be less likely to choose to re-sell their unused resource quota. This is seen in Eq. (\ref{sellersPayoffEqns}), as their payoffs are inversely proportional with respect to $\delta$, and in $q_s(p_r)=a\log(1+(1-\delta)p_r)$ (Eq. (\ref{eq:specificformQs})) where the sharing pool's supply $q_s$ is inversely proportional to $\delta$. This results in a lower quality/supply sharing pool, and therefore a smaller revenue which the MNO can obtain from sharing, making Region {\color{black}R2 (purely on-demand)} optimal at high $\delta$.
\textcolor{orange}{To summarize, it would have been expected that a high commission would give the platform a higher revenue, nevertheless our results show that it incurred a trade-off in dissuading reservation plan users from re-selling, hence decreasing the supply and revenue.}


\textcolor{red}{\textit{Impact of the on-demand price $p_o$ on the revenue and region :}} \textcolor{black}{In Fig. \ref{fig:combinedSimul}a, we show how the revenue varies as the on-demand price $p_o$ increases, for given values of $q_o,\delta=0.2, a=2, p_r=0.5$. As the price increases, the equilibrium region shifts from $R2$ (purely on-demand) to $R1$ (both on-demand and sharing) as more buyers choose the sharing platform due to the on-demand price increasing. As the on-demand price increases even further, the region transitions to $R3$ (purely sharing) due to the on-demand price being too high. Nevertheless, $R3$ is not optimal for the platform because it only collects a proportion $\delta$ of the amount transacted.}

\textcolor{red}{\textit{Impact of the normalization parameter $a$ on the optimal revenue and region :}} \textcolor{brown}{Given that it is not clear what is an appropriate normalization parameter which relates the supply function $q_s(p_r)$ to the supply of the on-demand pool $q_o$, we explore different normalization parameters. In contrast to Fig. \ref{fig:impactDelta}a, we set a smaller normalization parameter $a=1.5$ for Fig. \ref{fig:combinedSimul}b.
\textcolor{red}{Once again, to obtain the optimal revenue and region, we optimize over the prices $(p_o, p_r)$, given each combination of $($on-demand supply $q_o,$ commission $\delta,$ and normalization parameter $a)$.}
With a smaller $a$, a higher price $p_r$ comes with a lower quality/supply $q_s(p_r)$ relative to $q_o$ and therefore a lower payoff for buyers at the sharing pool. Hence, buyers have lower incentive to buy resources from the sharing pool.} That is why region R2 (purely on-demand) gives the optimal revenue for all values of $q_o$ in Fig. \ref{fig:combinedSimul}b. \textcolor{red}{Nevertheless, although R1's optimal revenue is not the optimal over all regions under this set of parameters, users will still buy from the sharing pool and it's revenue is non-zero, as seen in Fig. \ref{fig:combinedSimul}b.}

\textcolor{red}{\textit{Analyzing individual buyer payoffs :}} {\color{black}Finally,}
we plot individual buyer payoffs (Eq. (\ref{eq:BuyerPayoff})) at equilibrium conditions in Fig. \ref{fig:combinedSimul}c. As we can see, buyers with lower willingness to pay $u_i$ choose neither pool. As their willingness to pay (utility of the job computation) increases, their payoff increases. We can see that for the $q_o=0.1$ and $q_o=0.2$ plots, users with high willingness to pay choose the sharing pool, as visualised through the different gradients.

In summary, our simulation results show that under the combination of low $\delta$, low $q_o$, and large $a$, having the sharing platform (Region R1) is optimal.
Besides, it is not optimal to have Region R3 (purely sharing), due to the MNO only being able to take a proportion $\delta$ of the revenue $N_{\text{share }} p_r \delta (1-\frac{p_r}{q_s(p_r)})$.

\section{Prototype and Preliminary User Studies}
\label{sec:prototype}
\subsection{Prototype Description}
\begin{figure}
\centering
\includegraphics[
angle=0,scale=0.22]{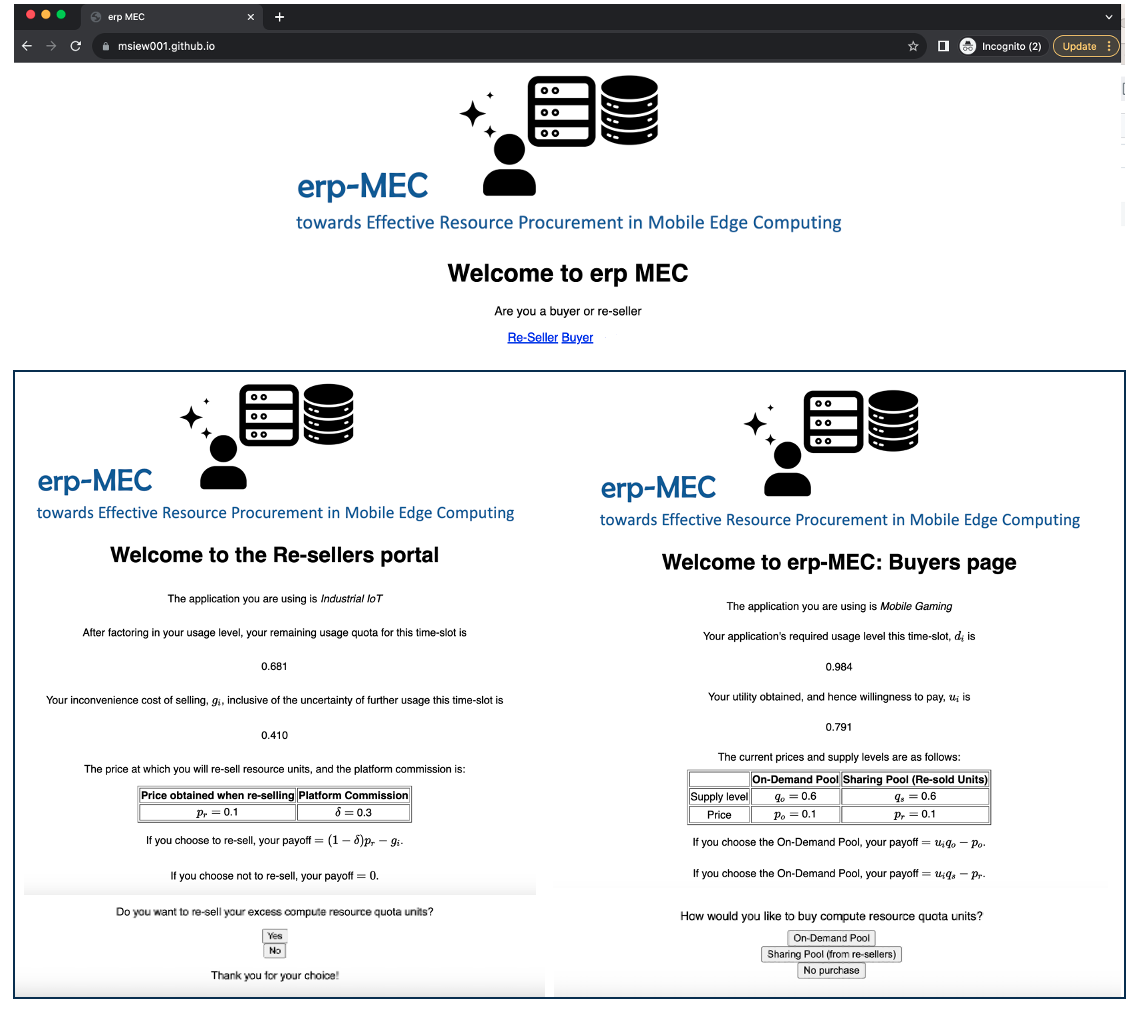}
\caption{\textcolor{violet}{Screnshots of our prototype - an interactive website, which can capture user decision making given different parameter settings, for user studies. It's url is https://shikhsh.github.io/erpMEC/.}}
\label{fig:websiteView}
\end{figure}

\begin{figure}
\centering
\includegraphics[
angle=0,scale=0.158]{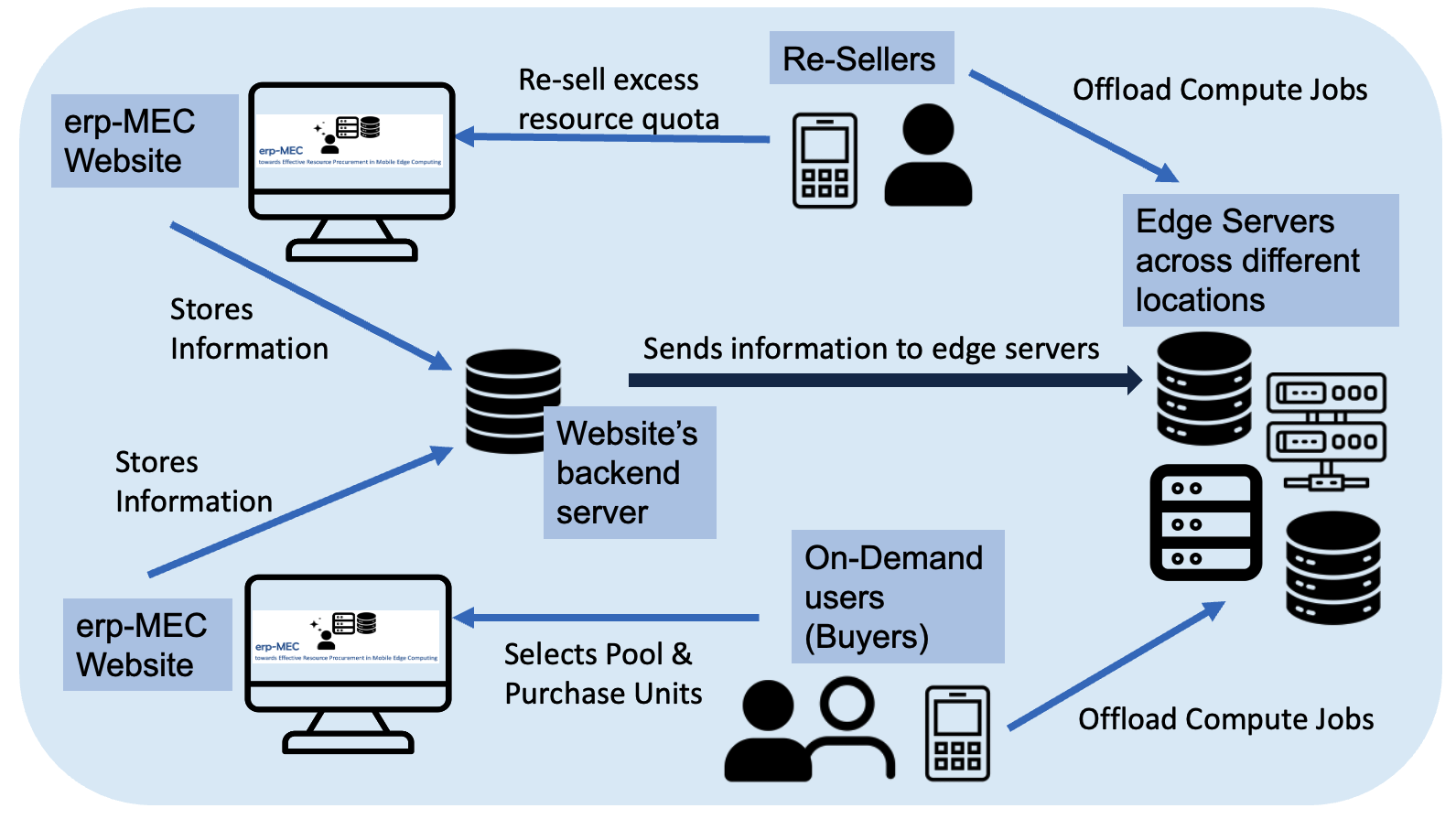}
\caption{\textcolor{violet}{How the real-world version of our prototype is integrated with the rest of the MEC system.}}
\label{fig:prototype_sysModel}
\end{figure}

\textcolor{violet}{We created a prototype, an interactive website (see Fig. \ref{fig:websiteView} for screenshots), which is able to capture user decisions given the different parameters. 
The interactive website's url is https://shikhsh.github.io/erpMEC. In Fig. \ref{fig:prototype_sysModel} we illustrate how the real-world version of our prototype integrates with the rest of the MEC system.}

\setlength{\textfloatsep} {3pt}

\textcolor{violet}{In our prototype, when users enter the website, they will be given a choice to select whether they are a buyer or re-seller.
In the re-sellers portal, the user is shown 1) their specific edge computing application, randomly sampled from the list ['IoT Sensing Analytics', 'Distributed AI', 'Predictive Maintence', 'Remote Monitoring', 'Industrial IoT', 'Augmented Reality Display'], 2) their remaining usage quota this time-slot, after factoring in their planned usage level, and 3) their inconvenience cost of re-selling, in light of potential further usage this time-slot. Note that the inconvenience cost also includes the cost of opening the website or app. 4) The re-selling price and platform commission, 5) the payoff equations under the different options, and finally 6) the \textbf{option buttons “re-sell” and “no”, for the user to choose from}.
Parameters 1, 2, and 3 are \textbf{user specific parameters}, and are uniformly randomly generated upon each user’s entrance for our prototype. When the number of users who use our prototype in user studies is large, we can compare the empirical distribution of user decisions to that of our model. The parameters in 4 are fixed for all users, and can be varied across user studies to view how users make decisions, and the aggregate outcome, under different settings. In practice, the information shown such as the type of application and the inconvenience cost (given potential further usage that time-slot) can be pulled from the user's usage history. The information on the user's remaining resource quota comes from the website's back-end calculations.}

\textcolor{violet}{
In the buyers‘ portal, the user is shown 1) their specific edge computing application, randomly sampled from the list ['IoT Sensing Analytics', 'Mobile Gaming', 'Augmented Reality', 'Data Analytics on Device', 'Smart Home IoT Analytics'] 2) their usage demand this time-slot, 3) their utility received, and hence their willingness to pay, 4) the supply and prices under the sharing pool and re-selling pool, 5) the payoff equations under the different options, and 6) \textbf{the option buttons “On-Demand Pool”, “Sharing Pool”, “No purchase”, for the user to choose from}.
As per the re-seller’s portal, parameters 1, 2, and 3 are \textbf{user specific parameters}, and are uniformly randomly generated upon each user’s entrance for our prototype. When the number of users who use our prototype in user studies is large, this models the equilibrium and aggregate outcome of heterogeneous users decision making. The parameters in 4 are fixed, and can be varied across user studies to view how users make decisions, and the aggregate outcome, under different settings. In practice, the information shown such as the type of application can be pulled from the user's usage history. The information on the user's remaining resource quota comes from the website's back-end calculations.
In practice, the users' willingness to pay will not be shown on the website, but is inherent in users' decision making.
}
\begin{figure}
\centering
\includegraphics[
angle=0,scale=0.55]{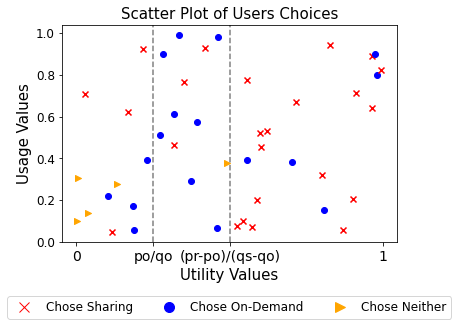}
\caption{\textcolor{violet}{Results of preliminary user study: The scatter points indicate the user choices given the utility and usage values they were shown on the website. The dotted lines indicate the region boundaries from the analytical model in the paper.}}
\label{fig:buyersScatterPlot}
\end{figure}
\setlength{\textfloatsep} {3pt}

\begin{table*}[t]
    \centering
    \begin{tabular}{|c|c|c|c|}
        \hline
          Region & \textbf{Sharing} & \textbf{On Demand} & \textbf{Neither} \\
        \hline
        Analytical Model (Prop 1-2) & $\frac{pr-po}{qs-qo}<u_i<1$ & $\frac{po}{qo}<u_i<\frac{pr-po}{qs-qo}$ & $u_i<\frac{po}{qo}$ \\
        \hline
        Percentage agreement of data and model & 77.3$\%$ & 66.7 $\%$ & 33.3 $\%$ \\
        \hline
        Test Statistic (of KS test) & 0.2917 & 0.2941 & 0.2667\\
        \hline
        P-value (of KS test) & 0.2406 & 0.6487 & 0.9629 \\
        \hline
        Are the data and & No significant evidence & No significant evidence & No significant evidence \\
        model distributions distinct? & & & \\
        \hline
    \end{tabular}
    \caption{\textcolor{violet}{Results for Buyers: Comparing data from user studies with the expected values from the model for each choice: Sharing, On-demand and Neither, through calculating the percentage agreement, and through using the Kolmogorov-Smirnov (KS) test. Our results show that there is no significant evidence of the data following a distinct distribution from our paper's model (at significance level 0.05).}}
    \label{tab:buyersResults}
\end{table*}

\begin{table*}[t]
    \centering
    \begin{tabular}{|c|c|c|}
        \hline
          Region & \textbf{Re-selling} & \textbf{Not re-selling}   \\
        \hline
          Analytical Model (Eq 1) & $g_i < (1-\delta)p_r$ & $g_i > (1-\delta)p_r$   \\
        \hline
        Percentage agreement with model (within region) & 90.9$\%$ & 52.9$\%$ \\
        \hline
        Test Statistic (of KS test) & 0.4444 & 0.1824 \\
        \hline
        P-value (of KS test) & 0.0975 & 0.9532  \\
        \hline
        Are the data and & No significant evidence & No significant evidence \\
        model distributions distinct? & & \\
        \hline
    \end{tabular}
    \caption{\textcolor{violet}{Results for Re-sellers: Comparing data from user studies with the expected values from the model for each choice: Re-selling, and not re-selling, through calculating the percentage agreement, and through using the Kolmogorov-Smirnov (KS) test. Our results show that there is no significant evidence of the data following a distinct distribution from our paper's model (at significance level 0.05).}}
    \label{tab:resellersResults}
\end{table*}

\textcolor{violet}{ 
Our prototype \textbf{stores a string of information upon each user click, on our back-end server} (Fig \ref{fig:prototype_sysModel}). For re-sellers, the string contains [remaining usage level for the time-slot, inconvenience cost/ uncertainty of usage $g_i$, re-selling price $p_r$, commission level $\delta$, type of application, Y or N indicating the \textbf{user's choice of whether or not to re-sell}, the time stamp].
For buyers, the string contains [the user's usage requirement, the utility/ willingness to pay $u_i$, the on-demand supply $q_o$, the sharing supply $q_s$, the on-demand price $p_o$, the sharing price $p_r$, \textbf{the user's choice (Sharing, On-demand, No purchase)}, the time stamp].
Our prototype stores these information, without storing the identity or any other information regarding the user. All clicks are anonymous. 
In practice, the back-end server will relay the decisions of the users to the edge. Based on this, the resource units at edge servers will be correspondingly allocated to the different users, by the computer at the edge server (Fig \ref{fig:prototype_sysModel}).}

\subsection{Preliminary User Study}
\textcolor{violet}{The goal of the user study is to verify how accurately our decision model (Eq. 1 for re-sellers, and Prop. 1-2, Fig 3a for buyers) models the way users make decisions. We will evaluate how similar the empirical distribution function of the \textit{data from the user study}, and the empirical distribution function of the \textit{data from the model's predictions}, are. }

\textcolor{violet}{For this preliminary user study, we set the sharing supply available to be slightly higher than the on-demand supply ($q_o=0.6, q_s=0.7$), and the sharing price to be slightly higher than the on-demand price ($p_o=0.15, p_r=0.2$). For our participants, we have had 44 participants be buyers, and 28 participants be re-sellers. 
Our results are in Fig. \ref{fig:buyersScatterPlot}, Tables \ref{tab:resellersResults} and \ref{tab:resellersResults}. Fig \ref{fig:buyersScatterPlot} is a scatter plot of buyers clicks, given the usage and utility values they were shown on the website. Under our paper's math model (Proposition 2/ Fig 3a of paper), buyers will choose the sharing pool when their utility value $\frac{pr-po}{qs-qo}<u_i<1$, will chose the on-demand pool when $\frac{po}{qo}<u_i<\frac{pr-po}{qs-qo}$, and will choose neither when $u_i<\frac{po}{qo}$. We can see in the scatter plot that the on-demand and sharing region largely agrees with the paper's model, which is corroborated by the percentage agreement with model in Table \ref{tab:buyersResults}. 
The 
\textit{percentage agreement} is calculated as follows: For each range of utilities $u_i$ corresponding to the regions [sharing, on-demand, neither] (i.e. [$\frac{pr-po}{qs-qo}<u_i<1$, $\frac{po}{qo}<u_i<\frac{pr-po}{qs-qo}$, $u_i<\frac{po}{qo}$ ]), we will calculate the percentage of data points from our user study which agree with our model. For e.g., in the "sharing" region, there are altogether 22 data points, and 17 agree with our model (Prop 1-2, and Fig. 3a), giving a percentage of $77.3\%$.
We note that there is randomness inherent in user clicks (decisions), as a math model can not fully capture user behavior. Furthermore decisions in a prototype will have differences from user behaviors in the actual system. 
Next we perform the Kolmogorov-Smirnov (KS) test for both the buyers and re-sellers data. We obtain the \textbf{model's data samples} through using the given utility values, and obtaining the corresponding decisions according to the model (Prop 1-2, Fig 3a for buyers, and Eq 1 for re-sellers). With this and our data samples, we perform a KS test to evaluate whether, for each region (Sharing, On Demand, Neither, for buyers, Re-selling, Not Re-selling, for re-sellers) the two set of data points (data from user studies and model) can considered to be from the same distribution. We perform the KS test using the scipy library in python. The KS test calculates the mean and standard deviation for each set of data points, and obtains the KS test test statistic, 
\begin{equation}
    D=\sup_x|F_{data}(x)-F_{model}(x)| 
\end{equation}
which quantifies the distance between the empirical distribution functions of the two data samples (model and user studies data).
We choose a confidence level of $95\%$.
As the p-values are above the significance level of $0.05$, there is no significant evidence that the data and model distributions are different.
Beyond our preliminary study in this work, further studies under different parameter settings, can be made with the help of our interactive prototype.
}

\section{Future Work}
\label{sec:futurework}

For future work, we will study the long-term scenario where users may switch between reserving resources in advance, and buying on the go through the on-demand and sharing pools. Here, another layer would be added to the Stackelberg game and users will make decisions based on their expected payoffs in the long run. The expected payoff would depend on the users' application requirements: the probability of them needing job computation at the edge consistently over all timeslots vs one-off and sporadically, the prices set by the MNO and the expected supplies of the different options, which is influenced by the aggregate decisions of other users.

Another line of investigation is to study how the sharing framework can be implemented across multiple edge nodes, with different users being associated with different nodes due to proximity. 
In light of this, the on-the-go (on-demand and sharing) queues and the sharing supply would differ across nodes. If users are allowed to associate with other nearby nodes, it is thus of interest to investigate how the MNO can jointly price and allocate the resources across nodes to the different users, to maximize revenue and minimize the number of unmet jobs. \textcolor{blue}{In this new study on incentives, a Stackelberg game formulation can be used to analyze and characterise the interaction between the different parties, along with a matching algorithm which matches users to resources across edge nodes, with the transmission cost and hence distance between user and node being a factor. In the new Stackelberg formulation, the user decision making functions will be modified to reflect the potential supply levels at different nodes, and the likelihood of the user attaining resources from different nodes, which is a function of the distance.}

\section{Conclusion}
\label{sec:conclusion}
In this paper
we {\color{black}proposed} a novel {\color{black}resource sharing platform} to enhance the usage of the on-demand {\color{black}pricing and resource reservation} models in edge computing. This helps to optimize the scenario where the on-demand users face a limited supply remaining at the edge server, whilst users under the {\color{black}resource reservation} model have reserved excess computing resources, by allowing the re-selling of unused resource quota.
We {\color{black}formulated} a {\color{black}3-stage} Stackelberg Game {\color{black} to investigate the user's optimal strategies and equilibrium behavior, as well as whether the MNO has an incentive to provide such a resource sharing platform. By analyzing the formulated game, we} characterised the optimal strategies of the re-sellers, buyers and MNO, and showed that there would be 4 regions (both sharing and on-demand, only on-demand, only sharing, none at all) at equilibrium after buyers' decisions.
Based on these 4 regions, we {\color{black} optimized} the MNO's revenue and analytically {\color{black}characterised} the MNO's optimal on-demand prices. 
Our numerical {\color{black}results show} that having a sharing pool gives the MNO an optimal revenue when the on-demand pool's supply is low and when the platform's commission is low. 

\section*{Acknowledgment}
This paper is supported in part by the National Research Foundation, Singapore and Infocomm Media Development Authority under its Future Communications Research $\&$ Development Programme, 
in part by the SUTD Presidential Postdoctoral Fellowship, under the Singapore MOE START Scheme, 
in part by NSF CNS-1751075, 
in part by SongShan Laboratory Foundation, under Grant No. YYJC022022007, in part by the open research fund of National Mobile Communications Research Laboratory, Southeast University, under Grant No. 2023D04, in part by the National Natural Science Foundation of China, under Grant No. 62301222.

\bibliographystyle{IEEEtran}

\bibliography{references} 

\ifCLASSOPTIONcaptionsoff
  \newpage
\fi



%


\begin{IEEEbiography}
[{\includegraphics[width=1in,height=1.25in,keepaspectratio]{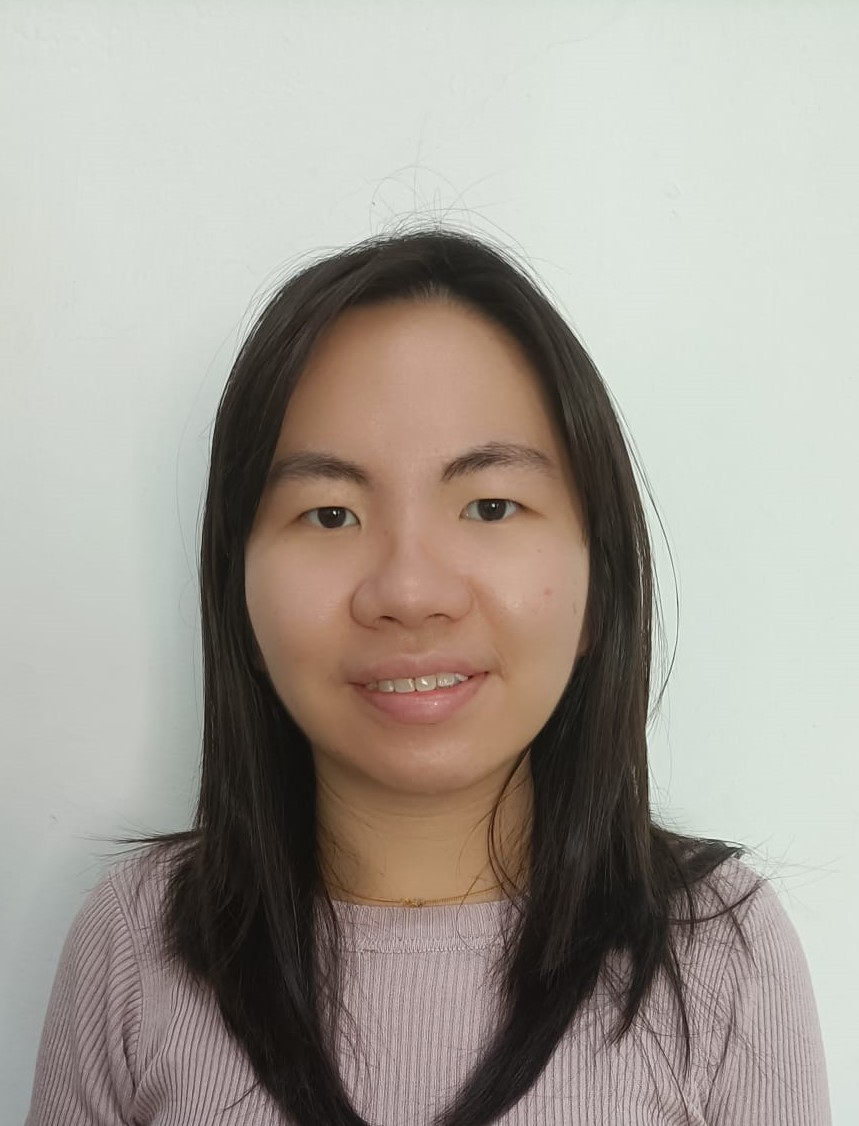}}]
{Marie Siew}(S'19, M'21) is currently a post-doctoral researcher at the Electrical and Computer Engineering department, Carnegie Mellon University. She received the B.Sc. degree in Mathematical Sciences from Nanyang Technological University, Singapore in 2016, and the Ph.D. degree from the Singapore University of Technology and Design in 2021. Her research interests are in the areas of network economics, edge computing, edge intelligence, and federated learning. 

Dr Siew received the Best Poster Award at ACM/IEEE IPSN 2023, the SUTD Presidential Postdoctoral Fellowship, under the Singapore Teaching and Academic Research Talent Scheme in 2021, the A*Star Computing and Information Science Scholarship in 2019, the A*Star Undergraduate Scholarship in 2012 from the Agency for Science, Technology and Research, and was in the CN Yang Scholars' Programme in NTU from 2012 to 2016.
\end{IEEEbiography}

\begin{IEEEbiography}
[{\includegraphics[width=1in,height=1.25in,keepaspectratio]{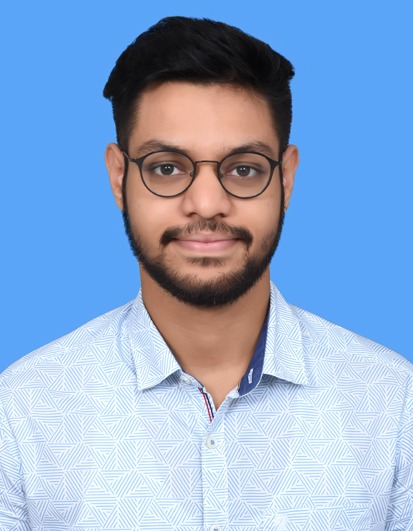}}]
{Shikhar Sharma} is a Masters Student in Electrical and Computer Engineering, Carnegie Mellon University. He received his Bachelors degree in 
Birla Institute of Technology and Science, Pilani in 2019. and worked as a software development engineer at Paypal from 2019-2021.
\end{IEEEbiography}

\begin{IEEEbiography}[{\includegraphics[width=1in,height=1.25in,keepaspectratio]{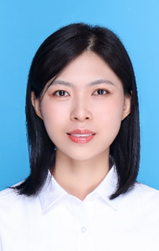}}]
{Kun Guo} (Member, IEEE) received the B.E. degree in Telecommunications Engineering from Xidian University, Xi'an, China, in 2012, where she received the Ph.D. degree in communication and information systems in 2019. From 2019 to 2021, she was a Post-Doctoral Research Fellow with the Singapore University of Technology and Design (SUTD), Singapore. Currently, she is a Zijiang Young Scholar with the School of Communications and Electronics Engineering at East China Normal University, Shanghai, China. Her research interests include wireless edge computing, caching, and intelligence.

\end{IEEEbiography}

\begin{IEEEbiography}
[{\includegraphics[width=1in,height=1.25in,keepaspectratio]{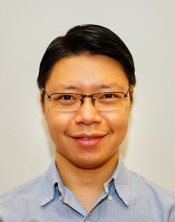}}]
{Desmond Cai}(S'10-M'16)  received the B.Sc. degree in electrical and computer engineering from Cornell University, Ithaca, NY, USA, in 2009 and the Ph.D. degree in electrical engineering from the California Institute of Technology, Pasadena, CA, USA, in 2016. 
He was a Research Scientist at IBM from 2019 to 2020, and a Scientist at the Institute of High Performance Computing from 2016 to 2019. His research interests include artificial intelligence, optimization, network economics, and game theory. 

Dr. Cai received the National Science Scholarship from the Agency for Science, Technology, and Research in 2004, the John G. Pertsch Prize from Cornell University in 2008, and the Sibley Prize from Cornell University in 2009. 
\end{IEEEbiography}

\begin{IEEEbiography}[{\includegraphics[width=1in,height=1.25in,keepaspectratio]{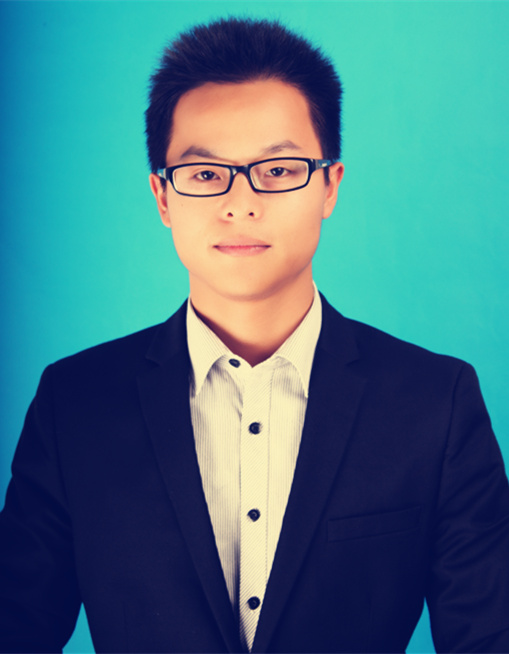}}]
{Wanli Wen} (S'15--M'20)  received the B.S. degree from Anhui University of Finance and Economics, Bengbu, China, in 2011, the M.S. degree in Communication and Information Systems from Hangzhou Dianzi University, Hangzhou, China, in 2014, and the Ph.D. degree in information and communications engineering from Southeast University, Nanjing, in 2019. From 2019 to 2020, he was a postdoctoral research fellow at Singapore University of Technology and Design. Since 2021, he has been an Assistant Professor in the School of Microelectronics and Communication Engineering, Chongqing University, China. His research interests include green communications, mobile edge computing and caching, and federated learning.
\end{IEEEbiography}

\begin{IEEEbiography}
[{\includegraphics[width=1in,height=1.25in,keepaspectratio]{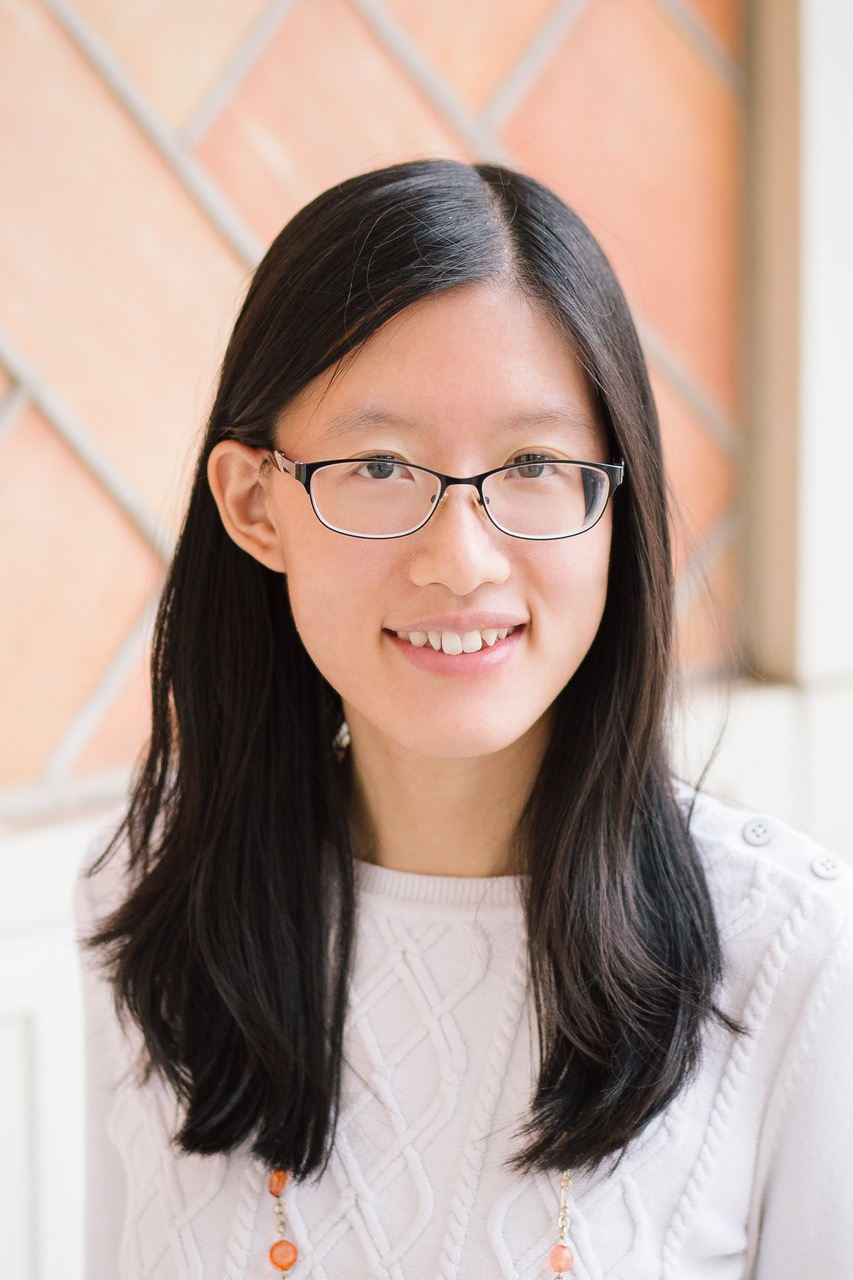}}]
{Carlee Joe-Wong} is the Robert E. Doherty Associate Professor of Electrical and Computer Engineering at Carnegie Mellon University. She received her A.B. degree (\emph{magna cum laude}) in Mathematics, and M.A. and Ph.D. degrees in Applied and Computational Mathematics, from Princeton University in 2011, 2013, and 2016, respectively. Her research interests lie in optimizing various types of networked systems, including applications of machine learning and pricing to cloud computing, mobile/wireless networks, and transportation networks. From 2013 to 2014, she was the Director of Advanced Research at DataMi, a startup she co-founded from her research on mobile data pricing. She received the NSF CAREER award in 2018 and the ARO Young Investigator award in 2019.
\end{IEEEbiography}

\begin{IEEEbiography}[{\includegraphics[width=1in,height=1.25in,keepaspectratio]{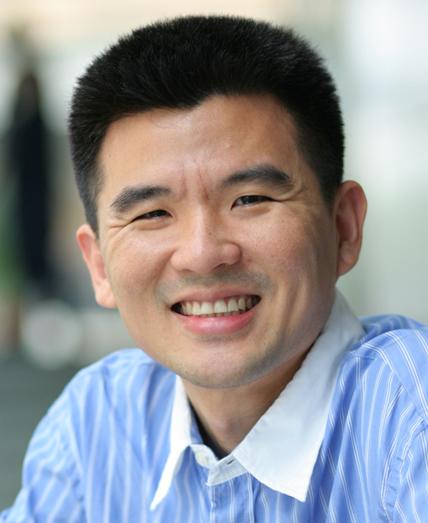}}]
{Tony Q.S. Quek}(S'98-M'08-SM'12-F'18) received the B.E.\ and M.E.\ degrees in electrical and electronics engineering from the Tokyo Institute of Technology in 1998 and 2000, respectively, and the Ph.D.\ degree in electrical engineering and computer science from the Massachusetts Institute of Technology in 2008. Currently, he is the Cheng Tsang Man Chair Professor with Singapore University of Technology and Design (SUTD) and ST Engineering Distinguished Professor. He also serves as the Director of the Future Communications R\&D Programme, the Head of ISTD Pillar, and the Deputy Director of the SUTD-ZJU IDEA. His current research topics include wireless communications and networking, network intelligence, non-terrestrial networks, open radio access network, and 6G.

Dr.\ Quek has been actively involved in organizing and chairing sessions, and has served as a member of the Technical Program Committee as well as symposium chairs in a number of international conferences. He is currently serving as an Area Editor for the {\scshape IEEE Transactions on Wireless Communications}. 

Dr.\ Quek was honored with the 2008 Philip Yeo Prize for Outstanding Achievement in Research, the 2012 IEEE William R. Bennett Prize, the 2015 SUTD Outstanding Education Awards -- Excellence in Research, the 2016 IEEE Signal Processing Society Young Author Best Paper Award, the 2017 CTTC Early Achievement Award, the 2017 IEEE ComSoc AP Outstanding Paper Award, the 2020 IEEE Communications Society Young Author Best Paper Award, the 2020 IEEE Stephen O. Rice Prize, the 2020 Nokia Visiting Professor, and the 2022 IEEE Signal Processing Society Best Paper Award. He is a Fellow of IEEE and a Fellow of the Academy of Engineering Singapore.
\end{IEEEbiography}

%





\end{document}